\documentclass[aps,prb,twocolumn,superscriptaddress]{revtex4-1}

\usepackage{subfigure}
\usepackage{graphicx}
\usepackage{vmargin}
\usepackage[dvips]{color}
\usepackage{amsmath}
\usepackage{appendix}
\usepackage[mathscr]{euscript}
\usepackage{cancel}

\setmarginsrb{1in}{1.5in}{1in}{0.5in}{0in}{0.2in}{0in}{0in}

\newcommand{\de}{\text{d}}

\newcommand{\f}{\frac} 

\newcommand{\fo}[1]{\frac{1}{#1}}

\newcommand\Real{\mathfrak{Re}}
\newcommand\Ima{\mathfrak{Im}}

\newcommand\ve{\varepsilon}

\newcommand\bk{{\mathbf k}}

\newcommand\sH{{\mathscr H}}

\newcommand{\ket}[1]{\left| {#1} \right\rangle} 
\newcommand{\bra}[1]{\left\langle {#1} \right|}

\newcommand{\leri}[1]{\left( #1 \right)}
\newcommand{\ppar}[1]{\left| #1 \right|}
\newcommand{\psqrt}[1]{\left[ #1 \right]}
\newcommand{\pbk}[1]{\left\{ #1 \right\}}

\newcommand{\under}[2]{\textcolor{black}{\underbrace{\textcolor{black}{#1}}_{
\textcolor{black}{#2}}}}

\newcommand{\ct}[2]{\textcolor{black}{\cancelto{#1}{\textcolor{black}{#2}}}}

\begin{document}
\title{Cotunneling signatures of Spin-Electric coupling in frustrated 
triangular molecular magnets}

\author{J.F. Nossa}
\affiliation{Department of Physics and Electrical Engineering, Linnaeus 
University, SE-39182 Kalmar, Sweden}
\affiliation{Solid State Physics/The Nanometer Structure Consortium, Lund 
University, Box 118, SE-221 00 Lund Sweden}
\author{C.M. Canali}
\affiliation{Department of Physics and Electrical Engineering, Linnaeus 
University, SE-39182 Kalmar, Sweden}
\date{\today}

\begin{abstract}
The ground state of frustrated (antiferromagnetic) triangular molecular magnets 
is characterized by two total-spin $S =1/2$ doublets with opposite chirality. 
According to a group theory analysis [M. Trif \textit{et al.}, Phys. Rev. Lett. 
\textbf{101}, 217201 (2008)] an external electric field can efficiently couple 
these two chiral spin states, even when the spin-orbit interaction (SOI) is 
absent. The strength of this coupling, $d$, is determined by an off-diagonal 
matrix element of the dipole operator, which can be calculated by 
\textit{ab-initio} methods [M. F. Islam \textit{et al.}, Phys.~Rev.~B 
\textbf{82}, 155446 (2010)]. In this work we propose that Coulomb-blockade 
transport experiments in the cotunneling regime can provide a direct way to 
determine the spin-electric coupling strength. Indeed, an electric field 
generates a $d$-dependent splitting of the ground state manifold, which can be 
detected in the inelastic cotunneling conductance. Our theoretical analysis is 
supported by master-equation calculations of quantum transport in the 
cotunneling regime. We employ a Hubbard-model approach to elucidate the 
relationship between the Hubbard parameters $t$ and $U$, and the spin-electric 
coupling constant $d$. This allows us to predict the regime in which the 
coupling constant $d$ can be extracted from experiment.
\end{abstract}

\pacs{75.50.Xx, 75.75.-c, 73.23.-b}

\maketitle

%
%
\section{Introduction}
\label{sec:Introduction}

Molecular magnets (MMs)\cite{Gatteschi2006} represent a rich playground for exploring quantum mechanics at 
the
nanoscale, and are intensively investigated 
both in condensed matter physics
and chemistry. MMs,
rationally designed and realized by chemical engineering,\cite{Wedge2012} are promising building blocks of 
electronic devices for molecular spintronics,\cite{Bogani2008,Sanvito2011} and for classical\cite{Affronte2009} 
and quantum information processing.\cite{Leuenberger2001, Lehmann2007, Ardavan2007}
For applications in quantum computation, MMs with frustrated antiferromagnetic coupling between spins are 
particularly 
promising, since at low energies they behave effectively as magnetic two-level systems with long spin coherent 
times,
which can be used as qbits to
encode and manipulate quantum information.\cite{Ardavan2007, Wedge2012} 
One outstanding issue in quantum information processing is the need of 
realizing fast control and switching between
quantum spin states. Standard spin-control techniques such as electron
spin resonance (ESR), carried out by time dependent magnetic fields, have limitations, since in practice it is 
difficult to 
achieve 
switching times of the order of nanoseconds for large enough fields. The need to achieve spatial resolutions of 
the order
of 1 nm represents another serious challenge for spin manipulations via magnetic fields. For these reasons,
control via electric fields seems to be a much more 
promising alternative, since strong electric fields 
can be switched on and off fast, and
applied selectively to nanoscale regions.\cite{Andre2006, Hirjibehedin2006, Bleszynski-Jayich2008}
  
Electric control and manipulation of magnetic properties is an important topic in solid state physics, 
presently studied in multiferroic materials, dilute
magnetic semiconductors and topological insulators. The electric control of nanomagnets presents both hard 
challenges and novel
possibilities. Since electric fields do not couple directly to spins, electric control can typically occur only 
indirectly, 
e.g., via a
manipulation of the spin-orbit interaction (SOI). Indeed, interesting spin-electric effects induced solely by 
SOI have been 
realized
in semiconductor quantum dots.\cite{kowack2007} The applicability of this procedure in MMs on the other hand is 
much harder, 
since the relative strength of the SOI scales with the volume of the system, implying that impractically large 
electric fields 
are required for 
systems of the order of a few nanometers. 
Therefore alternative schemes for efficient spin-electric coupling in MMs have been proposed. One example 
relies on the electric
manipulation of the spin exchange 
constant\cite{sanvito_nat_mat_2009, van_der_zant_2010}
which can trigger various level crossings between magnetic states of a different total spin.
Here we are interested in another type of spin-electric coupling, made it possible 
in certain antiferromagnetic MMs by the lack of inversion symmetry, as proposed
by Trif {\it et al.}.\cite{Trif2008} It turns out that in some of these antiferromagnetic molecules, 
such as the triangular \{Cu$_3$\} and \{V$_3$\} MMs,\cite{Choi2006,Yamase2004} 
and other odd-spin rings, an electric field can couple spin states 
through a combination of exchange and
chirality of the spin-manifold ground state (GS). For triangular MMs this coupling is nonzero even in the 
absence SOI.

\begin{figure}[tbp]
	\centering
		\includegraphics[scale=0.14]{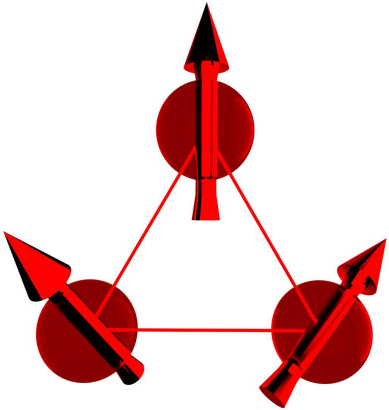}
	\caption{Schematic representation of a triangular molecule.}
	\label{fig:trianglespins}
\end{figure}

\begin{figure}[tbp]
	\centering
		\includegraphics[scale=0.19]{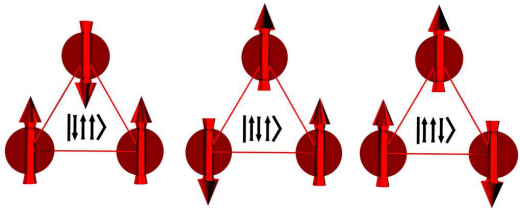}
	\caption{The three independent spin configurations associated with 
total spin projection $S_z=1/2$. The two chiral ground states $|E'_{\pm}, S_z =  1/2 
\rangle$ are linear combinations of these states.}
	\label{fig:phis}
\end{figure}

The low-energy physics of a triangular MM  can be described by  
three identical $1/2$-spin Cu cations, located at ${\bf r}_j,\; j = 1, 2, 3$, interacting via an
antiferromagnetic (Heisenberg) exchange coupling (see Fig.~\ref{fig:trianglespins}). 
%
%
%
%
The ground state consists of two total-spin $S = 1/2$ doublets,
$|E'_{\pm}, S_z = \pm 1/2 \rangle$, of opposite spin
chirality, $E'_{\pm} $, which are degenerate in the absence of spin-orbit interaction. (Here $E'$ refers to the
 two-dimensional 
irreducible representation (IR) of the $D_{3h} $ symmetry group of the triangular MM, spanned 
by the two states, $|E'_{+}, S_z \rangle$ and $|E'_{-}, S_z \rangle$.)
The states $|E'_{\pm}, S_z =  1/2 \rangle$ can be written as linear combinations of the three frustrated spin 
configurations 
shown in Fig.~\ref{fig:phis}.
%
%
%
%
%
%

According to an analysis based on group theory,\cite{Trif2008, Trif2010}
the matrix elements of the components of the operator 
${\bf R} = \sum_{j= 1}^3{\bf r}_j$ in the triangular molecule plane, 
$X_{\pm} = \pm X + i Y$, between states of opposite chirality, do not vanish
\begin{equation}
e \langle E'_+, S_z| X_-| E'_-,  S_z\rangle = e \langle E'_-, S_z| X_+| E'_+,  S_z\rangle = 2i d\,.
\label{dipole_me}
\end{equation}
In Eq.~(\ref{dipole_me}) $e$ is the electron charge, 
$i = \sqrt {-1}$, and the real number $d$ has the units 
of an electric dipole moment.
All the other matrix elements of $\bf R$ in the subspace
spanned by $\{|E'_{\pm}, S_z = \pm 1/2 \rangle\}$ are zero.
The nonzero value of $d$ is in fact related to the 
existence of a nonzero electric dipole moment in each of the three
frustrated spin configurations of Fig. ~\ref{fig:phis} that compose the chiral ground 
states.\cite{Bulaevskii2008,Khomskii2010, Islam2010, Khomskii2012}

An electric field $\pmb{\varepsilon}$  couples to the triangular MM via $e \pmb{\varepsilon} \cdot{\bf R}$. 
Then the non-zero matrix elements in Eq.~(\ref{dipole_me}) ensure that 
the amplitude of the spin-electric coupling 
between chiral states is {\it linear} in the field. 
Note that the electric-field--induced transitions conserve the total spin. 
However, in the presence of an
additional small dc magnetic field that mixes the spin states, this
spin-electric coupling can generate efficient electric
transitions from one spin state to another.  

The relevance of this spin-electric mechanism for
qubit manipulation and qbits coupling clearly depends 
on the value of the electric dipole moment $d$. It has been proposed\cite{Trif2008}
that 
an experimental estimate of $d$ can 
in principle be provided by ESR measurements in static
electric fields. Nuclear magnetic resonance, magnetization and specific heat measurements have also
been proposed to determine the strength of the coupling experimentally.\cite{Trif2010} 
As far as we know these measurements have not yet been performed.
Theoretically, a Hubbard model approach can provide
understanding and a rough estimate of $d$ in terms of a small number of 
Hubbard model parameters.\cite{Trif2010}

In practice, a microscopic evaluation of $d$ can only be provided by
first-principles calculations. In fact, in Ref.~\onlinecite{Islam2010} we
have carried out Density Functional Theory (DFT) studies of a
 \{Cu$_3$\} MM, and shown that $d$ is of the order of $e 10^{-4} a$, 
where $a$ is the
Cu-Cu separation. At electric fields of the order of $10^8$ V/m, easily accessible
in the vicinity of a scanning tunneling spectroscope (STM) tip, 
a $d$ of this size would ensure transition times of
the order of 1 ns. More recent DFT calculations\cite{Nossa_SE_V15_2013} have shown that the value of $d$
in other triangular molecules, such as \{V$_3$\}  \{Cu$_3O_3$\} and \{V$_{15}$\},  can be one or two orders of 
magnitude larger
than in \{Cu$_3$\}. 

In this paper we carry out a theoretical study of quantum
transport through an {\it individual}
triangular antiferromagnetic MM displaying the spin-electric coupling, arranged in a single-electron transistor 
(SET) 
geometry. 
The main motivation of this work is to
investigate whether the coherent coupling of the two spin chiral states induced and controlled by an electric 
field has
detectable consequences on the transport properties of the MM.
Our conclusion is that, in the cotunneling regime of Coulomb blockade transport, the GS energy splitting induced
by the electric field should be easily accessible and should provide a direct estimate of the strength of the
electric dipole moment parameter $d$. In this coherent regime, higher excited states of the MM could add as 
additional auxiliary
states that can be exploited to perform quantum gates.\cite{Trif2008} 
We also show that similar results could be obtained by
performing inelastic electron tunneling spectroscopy 
through the MM adsorbed on surface by means of STM techniques.  
For the modeling of the MM we use the Hubbard model 
approach introduced in Ref.~\onlinecite{Trif2010}.
This approach is quite convenient and transparent to
address the effect of an applied electric field on the molecular orbitals of the molecule leading to the spin-
electric coupling.
The parameters of the model are extracted from our previous first-principles calculations on \{Cu$_3$\}.
Quantum transport is studied by means of a quantum master equation including both sequential and cotunneling 
contributions.
Transport studies on triangular systems using a similar formalism 
have been done recently.\cite{Hoogdalem2013,Bulka2011,Weymann2011,Luczak2012,Hsieh2012,Xiong2012}
But our motivation is different and
an analysis of the spin-electric
effect in this system has not been considered so far. 

The paper is divided into the following.
four sections.
In Sec.~\ref{sec:Spinelectriccoupling}
we introduce a Hubbard approach to model the effect of the electric field
leading to the spin-electric coupling in terms of a few free
parameters.
In Sec.~\ref{sec:TransportModelandMethods} we introduce the model and the formalism to study
quantum transport and calculate the conductance in the sequential tunneling and cotunneling regime. In
Sec.~\ref{sec:Results} we present transport results. Finally, we summarize the
conclusions of our work in Sec. \ref{sec:Conclusions}

%
%
\section{Hubbard model approach to the spin-electric coupling}
\label{sec:Spinelectriccoupling}


In this section we introduce the Hubbard model approach developed in Ref.~\onlinecite{Trif2010} to analyze
the spin-electric coupling. This approach is very useful for three  reasons.  Firstly, it describes the
effect of the applied electric field on the orbital degrees of freedom of the molecular magnet (MM), and therefore it elucidates 
the emergence of the spin-electric coupling at the microscopic level. Secondly, it permits the description of
the spin-electric coupling in terms of a few parameters that can be evaluated by first-principles methods.
Last but not least, it provides the natural framework to study later on quantum transport.

Before we introduce the Hubbard model, it is convenient to summarize the main results of the spin-electric coupling
using the language of a spin Hamiltonian,\cite{Trif2008} in part already anticipated in the introduction, which will then 
emerge again
from the Hubbard model.

The ground-state manifold of a three-site spin $s= 1/2$ Heisenberg antiferromagnet, with isotropic exchange constant $J$,
is given by the two doubly-degenerate chiral doublets

\begin{equation}
|E'_{\pm }, S_z = \frac{1}{2}\rangle =\frac{1}{\sqrt{3}} \big (|\downarrow
\uparrow \uparrow\rangle + \epsilon_{\pm}
|\uparrow\downarrow\uparrow\rangle + \epsilon_{\mp}
|\uparrow\uparrow\downarrow\rangle\big ) \label{eq:chi_pm_sz+}\;,
\end{equation}
where $\epsilon_{\pm}= \exp\left(\pm 2\pi i / 3\right)$.
These states are eigenstates of the total spin $\bf S ^2$  with eigenvalue $S= 1/2$, and of the total $z$ component $S_z$,
with eigenvalue $1/2$.
The three spin configurations in Eq. (\ref{eq:chi_pm_sz+})  are shown in Fig.~\ref{fig:phis}). 
Similar linear combinations can be written
for the 2 eigenstates of $S_z$ with eigenvalue $1/2$.  These states are also eigenstates of the 
$z$-component of the chiral spin operator
\begin{equation}
C_z= \frac{4}{\sqrt{3}} {\mathbf s}_1\cdot {\mathbf s}_2 \times
{\mathbf s}_3 \label{eq:chiraloperator}\;,
\end{equation}
with eigenvalue $\pm 1$. (The $\pm$ in $E'_{\pm }$ refers to this quantum number.) 

The lowest excited state, separated from the ground state (GS) by an energy of order $J$, is the fourfold degenerate eigenstate 
of $\bf S ^2$, with eigenvalue $3/2$. The element of this quartet that is an eigenstate of
 $S_z$ with eigenvalue $1/2$, is written
in terms of the same three spin configurations of Fig.~\ref{fig:phis} as
\begin{equation}
|A'_1, S_z = \frac{1}{2}\rangle =\frac{1}{\sqrt{3}} \big (|\downarrow \uparrow \uparrow\rangle + 
|\uparrow\downarrow\uparrow\rangle +
|\uparrow\uparrow\downarrow\rangle\big )\label{eq:S=3/2_Sz=1/2}\;.
\end{equation}
The four states $|A'_1, S_z \rangle$ form four $A'_1$ one-dimensional IR of the 
symmetry group $D_{3h}$. Note that the expectation value of $C_z$ for the states $|A'_1, S_z \rangle$ vanishes.

The SOI-induced Dzyaloshinskii-Moriya (DM) interaction splits the chiral GS manifold into two two-dimensional subspaces.
As we discussed in the introduction, an electric field couples states of opposite chirality.
These two interactions can be represented  by the following low-energy effective spin Hamiltonian\cite{Trif2008} 
\begin{equation}
H^{\rm spin}_{\rm eff} = \Delta_{\rm SOI} C_z\,S_z 
+ d \pmb{\varepsilon} \cdot {\bf C}_{\parallel}
\label{H_spin_eff}
\end{equation}
where ${\bf C}_{\parallel} \equiv (C_x, C_y,0)$ is the  component of the chiral operator in the $xy$ plane.
In Eq.~(\ref{H_spin_eff}) the energy $\Delta_{\rm SOI}$ is proportional to the SOI coupling strength, and turns out to be
equivalent to the DM coupling constant $D$. The parameter $d$ is the 
electric dipole moment introduced in Eq.~(\ref{dipole_me}).  
We will now see how this effective spin Hamiltonian emerges from the Hubbard model approach.\cite{Trif2010} 

The second quantized one-band Hubbard Hamiltonian reads
\begin{equation}
H_U = -\sum_{i,j }\sum_{\scriptstyle  \alpha}
\Big\{ t_{ij} c_{i\alpha}^{\dagger}c_{j\alpha}^{\phantom{\dagger}}+
{\rm h. c.}\Big\} +\frac{1}{2}U\sum_i n_{i\uparrow}\, n_{i\downarrow}\;,
\label{eq:Hu}
\end{equation}
where $c_{i\alpha}^{\dagger}$ ($c_{i\alpha}$) creates
(destroys) an electron with spin $\alpha$ at  site $i$,
$n_{i\alpha}=c_{i\alpha}^{\dagger}c_{i\alpha}$  is the particle
number operator and $t_{ij}$ is a spin-independent hopping parameter.  More 
precisely, the index $i$ labels a Wannier function localized at site $i$.  The 
first term represents the kinetic energy describing electrons hopping between 
nearest-neighbor sites $i$ and $j$. For $D_{3h}$ symmetry this term is 
characterized by a hopping parameter $t_{ij} = t$. The second term is an 
on-site repulsion energy of strength $U$, which describe the energy cost 
associated with having two electrons of opposite spin on the same site. In this 
model the interaction energy between electrons which are not on the same site 
is completely neglected. The Hubbard model is the simplest model describing the 
fundamental competition between the kinetic energy and the
interaction energy of electrons on a lattice.

The  spin-orbit interaction in the Hubbard model is described by adding the 
following spin-dependent hopping
term~\cite{friedel_1964, kaplan_1982, bonesteel1992, Trif2010}
\begin{equation}
H_{\rm SOI}  = \sum_{i,j}\sum_{\scriptstyle  \alpha, \beta}
\Big\{ c_{i\alpha}^{\dagger}\Big(i\frac {{\bf P}_{ij}}{2}
\cdot {\pmb{ \sigma}}_{\alpha\beta}\;\Big) c_{j\beta}^{\phantom{\dagger}}+
{\rm h. c.}\Big\}\;,
\label{eq:hubbard_soi_gen}
\end{equation}
where $\pmb{\sigma}=\sigma_x \hat{x} + \sigma_y \hat{y} + \sigma_z \hat{z}$ is 
the vector of the three Pauli matrices. A commonly used notation  for the Pauli 
matrices is to write the  vector index $i$ in the superscript, and the matrix 
indices as subscripts, so that the element in row $\alpha$ and column $\beta$ 
of the $i$th Pauli matrix is $\sigma^i_{\alpha \beta}$, with $i=x,y,z$. 
Here the vector ${\bf P}_{ij}$ is proportional to the matrix element of $ 
{\pmb{ \nabla}} V\times {\bf p}$ between the orbital parts of
the Wannier functions at sites $i$ and $j$; $V$ is the one-electron potential 
and $\bf p$ is the momentum operator. Clearly the spin-orbit term has the form 
of a spin-dependent hopping, which is added to the usual spin-independent 
hopping proportional to $t$. In Eq. (\ref{eq:hubbard_soi_gen}), spin-orbit 
coupling induces a spin precession about ${\bf P}_{ij}$ when an 
electron hops from site $i$ to site $j$.
This form of the spin-orbit interaction is a special case of Moriya's hopping 
terms\cite{Moriya1960} in the limit that all but one orbital energy is taken to 
infinity,\cite{kaplan_1982} and it is consistent with our choice of a one-band 
Hubbard model. The $x$ and $y$ components of ${\bf P}_{ij}$ describe processes 
with different spin, and because of the $\alpha_v$ symmetry, ${\bf P}_{ij} = p 
{\bf e}_z$. Therefore, because of the symmetry of the molecule, the free Hubbard 
parameters are reduced to three, namely, $t$, $U$ and $p$.

The final expression of the Hamiltonian describing the electrons in a triangular 
molecule, including the spin-orbit interaction, is
%
%
%
\begin{eqnarray}
H_{U+{\rm SOI}}  &=& \sum_{\scriptstyle i,\alpha}
\Big\{ c_{i\alpha}^{\dagger}\big (-t + i\lambda_{\rm SOI}\alpha\big )
c_{i+1\alpha}^{\phantom{\dagger}}+
{\rm h. c.}\Big\} \nonumber \\
&& +  \sum_{\scriptstyle i, \alpha}  \left(  \epsilon_0 n_{i\alpha} + 
\frac{1}{2}U  n_{i\alpha}\, n_{i\bar{\alpha}}  \right),
\label{eq:hubbard_soi}
\end{eqnarray}
where $\lambda_{\rm SOI} \equiv p/2 = {\bf P}_{ij}/ 2 \cdot {\bf e}_z$ is the
spin-orbit parameter, $\epsilon_0$  is the on-site orbital energy, and 
$\bar{\alpha}=-\alpha$.

We want to treat the two hopping terms perturbatively on the same
footing, by doing an expansion around the atomic limit $t/U\;, \
\lambda_{\rm SOI}/U \to 0$. In many molecular magnets $t
\gg\lambda_{\rm SOI}$. This turns out to be the case also for
\{Cu$_3$\}.\cite{Nossa2012} In other molecules the two hopping parameters are 
of the
same order of magnitude.

We are interested in the half-filled regime. From second-order perturbation 
theory in $t/U$, an antiferromagnetic isotropic exchange term emerges and 
it splits the spin degeneracy of the low-energy
sector of the Hubbard model, which is defined by the singly-occupied states.

The perturbative method requires the definition of the unperturbed states being 
the one-electron states
\begin{equation}
\left| \phi_i^{\alpha} \right\rangle = c^{\dagger}_{i\alpha} \left| 0 
\right\rangle,
\label{eq:oes}
\end{equation}
singly-occupied three-electron states
\begin{equation}
\left| \psi_i^{\alpha} \right\rangle = \prod_{j=1}^3 c^{\dagger}_{j\alpha_j} 
\left| 000 \right\rangle =\prod_{j=1}^3 \left| \phi_j^{\alpha_j} \right\rangle ,
\label{eq:sotes}
\end{equation}
with  $\alpha_j=\alpha$ for $j\neq i$ and $\alpha_j=\bar{\alpha}$, for $j=i$. 
Finally the doubly-occupied three-electron states
\begin{equation}
\left| \psi_{ij}^{\alpha} \right\rangle = c^{\dagger}_{i\uparrow} 
c^{\dagger}_{i\downarrow} c^{\dagger}_{j\alpha}\left| 000 \right\rangle ,
\label{eq:dotes}
\end{equation}
with $i=1,2,3$ and $j\neq i$. Note that the states in Eqs. 
(\ref{eq:oes})-(\ref{eq:dotes}) are eigenstates of the Hamiltonian, Eq. 
(\ref{eq:hubbard_soi}),  only in the absence of the hopping and spin-orbit 
parameter and with energies $\epsilon_0$, $3\epsilon_0$ and $3\epsilon_0 + U$, 
respectively. These states are not yet symmetry adapted states of the $D_{3h}$ 
point group. Symmetry adapted states can be found using the the projector 
operator formalism.\cite{Trif2010,Boris2006} One-electron symmetry adapted 
states can be written as a linear combinations of one-electron states, Eq. 
(\ref{eq:oes}), 
\begin{equation}
\left| \Phi_{A'_1}^{\alpha} \right\rangle = \frac{1}{\sqrt{3}} \sum_{i=1}^3	
\left| \phi_i^{\alpha} \right\rangle ,
\label{eq:oesasA}
\end{equation}
and
\begin{equation}
\left| \Phi_{E'_{\pm}}^{\alpha} \right\rangle = \frac{1}{\sqrt{3}} \sum_{i=1}^3	
\epsilon_{1,2}^{i-1} \left| \phi_i^{\alpha} \right\rangle~\;,
\label{eq:oesasE}
\end{equation}
where $A'_1$ and $E'_{\pm}$ are one-dimensional and two-dimensional IR 
in the $D_{3h}$ point group, respectively, and 
$\epsilon^k=\exp\left((2 \pi i/3)^k \right)^{1,2}$ is a phase factor. 
The three-electron symmetry adapted states for singly-occupied magnetic centers 
can be written as
\begin{equation}
\left| \psi_{A'_{1}}^{1\alpha} \right\rangle = \frac{1}{\sqrt{3}} \sum_{i=1}^3	
 \left| \psi_i^{\alpha} \right\rangle ,
\label{eq:tesassoA}
\end{equation}
and
\begin{equation}
\left| \psi_{E'_{\pm}}^{1\alpha} \right\rangle = \frac{1}{\sqrt{3}} 
\sum_{i=1}^3	\epsilon_{1,2}^{i-1}  \left| \psi_i^{\alpha} \right\rangle ,
\label{eq:tesassoE}
\end{equation}

The states $|\psi_{E'_+}^{1\, \alpha}\rangle$ and $|\psi_{E'_-}^{1\, 
\alpha}\rangle$ have total spin $S=1/2$ and $z$-spin projection  $S_z  = \pm 
1/2$.  These states are formally identical to the  chiral states given in 
the Eq.~(\ref{eq:chi_pm_sz+}),  and  are eigenstates of the 
Hubbard Hamiltonian when $t= \lambda_{\rm SOI} =0$. The tunneling and spin orbit interaction (SOI) mix 
the singly-occupied 
and doubly-occupied states.  Symmetry properties of the $D_{3{\text{h}}}$ point group 
dictate that the tunneling and SOI terms in the Hubbard Hamiltonian transform 
as the irreducible (IR) $A_1'$. Therefore, only states transforming 
according to the same IR could be mixed. The 
first-order correction in $t/U$ and $\lambda_{\text{SOI}}/U$ is obtained by 
mixing in doubly-occupied states\cite{Trif2010}
\begin{eqnarray}
|\Phi_{E'_{\pm}}^{1\, \alpha}\rangle
&\equiv &|\psi_{E'_{\pm}}^{1\, \alpha}\rangle
+ 
\frac {({\epsilon^1_2} -1)(t\pm \alpha \lambda_{\rm SOI})}{\sqrt{2}U}
 |\psi_{E^{'1}_{\pm}}^{2\, \alpha}\rangle \nonumber \\
 &&
+ \frac {3{\epsilon^1_1} (t\pm \alpha \lambda_{\rm SOI})}{\sqrt{2}U}
 |\psi_{E^{'2}_{\pm}}^{2\, \alpha}\rangle\;,
\label{eq:chiral_u1st}
\end{eqnarray}
where
\begin{equation}
|\psi_{E^{'1}_{\pm}}^{2\, \alpha}\rangle= \frac{1}{\sqrt{6}}\sum_{i=1}^3 
\epsilon_{1,2}^{i-1}\left(
|\psi_{i1}^{\alpha}\rangle +|\psi_{i2}^{\alpha}\rangle
 \right)\;,
\end{equation}
and
\begin{equation}
|\psi_{E^{'2}_{\pm}}^{2\, \alpha}\rangle  = \frac{1}{\sqrt{6}}\sum_{i=1}^3 
\epsilon_{1,2}^{i-1}\left(
|\psi_{i1}^{\alpha}\rangle -  |\psi_{i2}^{\alpha}\rangle
 \right)\;,
\end{equation}
are three-electron symmetry adapted states for doubly-occupied magnetic centers.

In the small $t/U \;, \ \lambda_{\rm SOI}/U $ limit, we can resort
to a spin-only description of the low-energy physics of the system.
The ground state manifold (corresponding to the states in 
Eq.~(\ref{eq:chiral_u1st})) is given by the two chiral spin states 
of Eq.~(\ref{eq:chi_pm_sz+}). In this low-energy 
regime, the orbital states correspond to the singly-occupied  localized atomic 
orbitals. The lowest
energy states have total spin $S=1/2$ and chirality $C_z =\pm 1$.
Using the same perturbative procedure, we can construct approximate Hubbard model states
corresponding to the $S= 3/2$ excited-state quartet of Eq.~(\ref{eq:S=3/2_Sz=1/2}). 
To first order in $t/U$  and $\lambda_{\rm SOI}/U$ one obtains
\begin{equation}
|\Phi_{A'_{1}}^{1\, \alpha}\rangle = |\psi_{A'_{1}}^{1\, \alpha}\rangle
\label{quartet}
\end{equation}
The energy of the $S= 3/2$ quartet is $3 J/2$ higher in energy than the energy of the chiral GS doublets,
with $J \approx  4 t^2/U$.

%
%
%
%
\begin{figure}[tbp]
	\centering
		\includegraphics[scale=0.25]{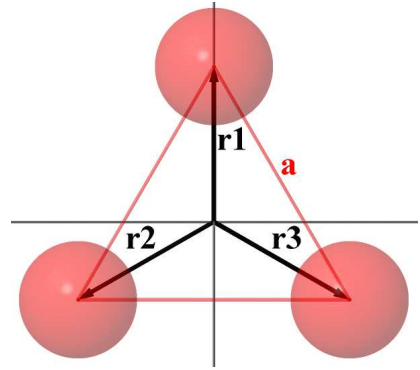}
	\caption{Coordinates of magnetic centers in a triangular molecule. 
${\bf r}_i$ is the coordinate of the $i$th electron.}
	\label{fig:triangle}
\end{figure}

We now introduce the effect of the external electric field. An external 
electric field $\pmb{\varepsilon}$ can couple to the molecule via two 
mechanisms. The first mechanism that we will study is by the modification  of 
the on-site energies $\epsilon_0$ via the Hamiltonian
\begin{equation}
H_{d-\varepsilon}^0 = \sum_{\alpha} \sum_{i=1}^3 \left( -e{\bf r}_i\cdot 
\pmb{\varepsilon} \right) c^{\dagger}_{i\alpha} c_{i\alpha} 
,
\label{eq:Hed0-1}
\end{equation}
where ${\bf r}_i$ is the coordinate vector of the $i$th magnetic center. From 
Fig.~\ref{fig:triangle}, the on-site electric Hamiltonian can be written as
\begin{eqnarray}
H_{d-\varepsilon}^0 &=& -ea \sum_{\alpha} \left[  
\frac{\varepsilon^y}{\sqrt{3}}  c^{\dagger}_{1\alpha} c_{1\alpha} 
-\frac{1}{2} \left( \varepsilon^x +\frac{\varepsilon^y}{\sqrt{3}}  \right) 
c^{\dagger}_{2\alpha} c_{2\alpha} 
\right. \nonumber \\
&& \left.+\frac{1}{2} \left( \varepsilon^x -\frac{\varepsilon^y}{\sqrt{3}}  
\right) c^{\dagger}_{3\alpha} c_{3\alpha}  \right] 
,
\label{eq:Hed0}
\end{eqnarray}
where $\varepsilon^{x,y}$ are the in-plane coordinates of the electric field, 
$e$ the electron charge and $a$ the distance between magnetic centers.

The second mechanism is given by the modification of the hopping parameters 
$t_{ii+1}$ and it can be written as
\begin{equation}
H_{d-\varepsilon}^1 = \sum_{\alpha} \sum_{i=1}^3 
t_{ii+1,\alpha}^{\pmb{\varepsilon}} c^{\dagger}_{i\alpha} c_{i+1\alpha} + 
\text{H.c.},
\label{eq:Hed1-1}
\end{equation}
where $t_{ii+1,\alpha}^{\pmb{\varepsilon}}=\left\langle \phi_{i}^{\alpha} 
|-e{\bf r}\cdot {\pmb{\varepsilon}} | \phi_{i+1}^{\alpha}\right\rangle$ are the 
modified  hopping parameters
due to the external electric field $\pmb{\varepsilon}$, $\phi_i^{\alpha}$  are 
the Wannier  states localized on the $i$th magnetic center with spin $\alpha$. 
These induced hopping parameters can be written as 
$t_{ii+1,\alpha}^{\pmb{\varepsilon}}=\sum_q q_{ii+1}^{\alpha}\varepsilon_q $, 
with $q_{ii+1}^{\alpha}=- e\left\langle \phi_{i}^{\alpha} | q | 
\phi_{i+1}^{\alpha}\right\rangle$ and $q=x,y,z$.
$D_{3h}$ point group symmetry properties, given by the dipole selection rules, 
reduce the number of free parameters induced by the electric field. Finding 
these free parameters is not an easy task when the basis set is composed of localized 
Wannier orbitals. In order to investigate the effect of the electric field on 
the triangular molecule, we switch from the localized Wannier basis set to the 
symmetry adapted basis set $\Gamma= A_1',E_{\pm}'$. Then we apply the 
transition dipole selection rules to the new induced hopping parameters. In the 
symmetry adapted states, the hopping-Hamiltonian, Eq. (\ref{eq:Hed1-1}), reads
\begin{equation}
H_{d-\varepsilon}^1 = \sum_{\alpha} \sum_{\Gamma \Gamma'}  
t_{\Gamma,\Gamma',\alpha}^{{\pmb{\varepsilon}}} c^{\dagger}_{\Gamma\alpha} 
c_{\Gamma'\alpha} + \text{H.c.},
\label{eq:Hed1-2}
\end{equation}
where $\Gamma,\Gamma'=A_1',E_+',E_-'$, 
$t_{\Gamma,\Gamma',\alpha}^{{\pmb{\varepsilon}}}=\sum_q 
q_{\Gamma\Gamma'}^{\alpha}E_q $, with $q=x,y,z$ and 
$q_{\Gamma\Gamma'}^{\alpha}=- e\left\langle \phi_{\Gamma}^{\alpha} | q | 
\phi_{\Gamma'}^{\alpha}\right\rangle$. Here $c^{\dagger}_{\Gamma\alpha} 
(c_{\Gamma\alpha})$ creates (destroys) an electron in the adapted state 
$\Gamma$ with spin $\alpha$. Note that in Eq. (\ref{eq:Hed1-2}) all the 
possible transitions are included, even those between states of the same 
symmetry adapted basis set. Dipole transition rules then will select the 
allowed transitions  and the corresponding states. Although symmetry properties 
control the dipole transition rules, they do not allow us to calculate the 
strength of the transitions. Detailed experimental measurements and/or accurate $ab$-initio 
calculations have to be carried out to determine them. In the $D_{3h}$ point 
group, the $(x,y)$ and $z$-coordinates span as the $E'$ and the $A'_1$ IR, 
respectively. We have grouped $x$ and $y$ because they form a 
degenerate pair within the $E'$ representation. From character tables of 
the $D_{3h}$ point group, the only allowed transitions correspond to 
\begin{eqnarray}
\left\langle \phi_{E_+'}^{\alpha} \right| x \left| 
\phi_{E_-'}^{\alpha}\right\rangle&=&-i 
\left\langle \phi_{E_+'}^{\alpha} \right| y \left| 
\phi_{E_-'}^{\alpha}\right\rangle \equiv -\frac{d_{EE}}{e}
\nonumber \\
\left\langle \phi_{A_1'}^{\alpha} \right| x \left| 
\phi_{E_+'}^{\alpha}\right\rangle&=&-i 
\left\langle \phi_{A_1'}^{\alpha} \right| y \left| 
\phi_{E_+'}^{\alpha}\right\rangle \equiv -\frac{d_{AE}}{e}
\label{eq:dEEpar}\\
\left\langle \phi_{A_1'}^{\alpha} \right| x \left| 
\phi_{E_-'}^{\alpha}\right\rangle&=&i 
\left\langle \phi_{A_1'}^{\alpha} \right| y \left| 
\phi_{E_-'}^{\alpha}\right\rangle \equiv -\frac{d_{AE}}{e}
\nonumber 
\end{eqnarray}
where $d_{EE}$ and $d_{AE}$ are the only two free parameters to be determined. 
Here we have used the symmetry rule that  the product $f_1 \otimes f_2 \otimes 
f_3 \neq 0$ if it spans the $A_1$ representation. All the other possible 
transitions are not allowed within the $D_{3h}$ symmetry group. Inserting these 
allowed transitions into the Hamiltonian, Eq. (\ref{eq:Hed1-2}), we have\cite{Trif2010} 
\begin{eqnarray}
H_{d-\varepsilon}^1 &=& \sum_{\alpha} \left[ d_{AE}\left(
 \bar{\mathcal{E}} c^{\dagger}_{A_1'\alpha} c_{E_-'\alpha}
+     \mathcal{E}  c^{\dagger}_{A_1'\alpha} c_{E_+'\alpha}      \right) \right. 
\nonumber \\
&&+ \left. d_{EE} 
 \bar{\mathcal{E}} c^{\dagger}_{E_-'\alpha} c_{E_+'\alpha}      \right]
  + \text{H.c.} 
,
\label{eq:Hed1}
\end{eqnarray}
where $\mathcal{E}=\varepsilon^{x}+i\varepsilon^{y}$ and  
$\bar{\mathcal{E}}=\varepsilon^{x}-i\varepsilon^{y}$. Note that the parameters 
$d_{AE}$ and $d_{EE}$  tell us about the possible dipole-electric transitions 
between states that span the $A_1'$-$E_{\pm}'$ and $E_+'$-$E_{-}'$ IR, 
respectively. From Eq. (\ref{eq:tesassoE}) we can see that the chiral states 
also span the $E_{\pm}$ IR. 

To take even more advantage of the symmetry of the triangular molecule, we now 
write the relationship between the second quantized operators $ 
c^{\dagger}_{i\alpha},c_{i\alpha}  $ and the symmetry adapted operators $ 
c^{\dagger}_{\Gamma\alpha},c_{\Gamma\alpha}  $. From Eqs. 
(\ref{eq:oes}),(\ref{eq:oesasA}) and (\ref{eq:oesasE}), we have 
\begin{equation}
\begin{pmatrix}
c^{\dagger}_{A_1'\alpha} \\
c^{\dagger}_{E_+'\alpha} \\
c^{\dagger}_{E_-'\alpha} 
\end{pmatrix}
=
\begin{pmatrix}
1 & 1 & 1 \\
1 & \epsilon & \epsilon^2 \\
1& \epsilon^2  &\epsilon
\end{pmatrix}
\begin{pmatrix}
c^{\dagger}_{1\alpha} \\
c^{\dagger}_{2\alpha} \\
c^{\dagger}_{3\alpha} 
\end{pmatrix}
,
\label{eq:l2as}
\end{equation}
where we have used $\epsilon^4=\epsilon$. From the last equation we can write the 
localized second quantized operators as a linear combination of symmetry adapted 
operators 
\begin{equation}
\begin{pmatrix}
c^{\dagger}_{1\alpha} \\
c^{\dagger}_{2\alpha} \\
c^{\dagger}_{3\alpha} 
\end{pmatrix}
=
\begin{pmatrix}
1 & 1 & 1 \\
1 & \epsilon^2 & \epsilon \\
1& \epsilon  &\epsilon^2
\end{pmatrix}
\begin{pmatrix}
c^{\dagger}_{A_1'\alpha} \\
c^{\dagger}_{E_+'\alpha} \\
c^{\dagger}_{E_-'\alpha} 
\end{pmatrix}
.
\label{eq:as2l}
\end{equation}

Now we can write the rest of the perturbed Hamiltonian, namely the 
$H_{d-\varepsilon}^0$ on-site electric field Hamiltonian (Eq. (\ref{eq:Hed0})) 
and $H_{\rm SOI}$ spin-orbit Hamiltonian (Eq. (\ref{eq:hubbard_soi_gen})), in 
terms of the symmetry adapted operators  
\begin{eqnarray}
H_{d-\varepsilon}^0&=&-\frac{i a e}{2\sqrt{3}} \sum_{\alpha}\left[
 \bar{\mathcal{E}} c_{E_+'\alpha}^{\dagger}c_{A_1'\alpha}
-  \mathcal{E}  c_{E_-'\alpha}^{\dagger}c_{A_1'\alpha}  \right.\nonumber \\
&&\left. \ \ \  \ \ \ \  \ \ \ \  \ \ \ \ 
+\bar{\mathcal{E}} c_{E_-'\alpha}^{\dagger}c_{E_+'\alpha}
    \right] + \text{ H.c.} \label{eq:AEed0}
,
\label{eq:Hed0as}
\end{eqnarray}
and 
\begin{equation}
H_{\rm SOI} = \sqrt{3} \lambda_{\rm SOI} \sum_{\alpha} \alpha
\left(
 c^{\dagger}_{E'_-\alpha} c_{E'_-\alpha} 
-c^{\dagger}_{E'_+{\bar \alpha}} c_{E'_+{\bar \alpha}} 
\right).
\label{eq:soias}
\end{equation}

We conclude this section with the following important considerations

\noindent
1. With the use of the symmetry properties of the triangular molecule, the Hubbard model in the
presence of SOI (Eq.~(\ref{eq:soias})) and an external electric field (Eqs.~(\ref{eq:Hed1}) and (\ref{eq:Hed0as})),
can be parametrized by five free parameters: 
$t$, $U$, 
$\lambda_{\rm SOI}$, $d_{EE}$ and $d_{AE}$. For a realistic molecular magnet, $t$, $U$, $\lambda_{\rm SOI}$ can be
extracted from first-principles calculations, as for example done in Ref.~\onlinecite{Nossa2012} for $\{ Cu_3 \}$. 
An analogous determination of the single-particle parameters $d_{EE}$ and $d_{AE}$ has not been attempted so far.
For localized orbitals, one expects $ea >> d_{EE}, d_{AE}$, and this the assumption that we will make in the paper.  

\noindent
2. Eqs.~(\ref{eq:Hed1}) and (\ref{eq:Hed0as}) and Eq.~(\ref{eq:soias}) are completely consistent with the effective
spin Hamiltonian result of Eq.~(\ref{H_spin_eff}), in that they imply a splitting of the chiral GS by the SOI, 
and a linear coupling of
the same states by an electric field. Note also that the SOI does not mix states of different chirality and/or spin.

\noindent
3. Clearly Eqs.~(\ref{eq:Hed1}) and (\ref{eq:Hed0as}) and Eq.~(\ref{eq:soias}) are single-particle Hamiltonian. In order to 
extract the electric-dipole moment $d$ and the DM splitting $\Delta_{\rm SOI}$ appearing in Eq.~(\ref{H_spin_eff}), one has to 
take matrix elements of these Hamiltonians between many-body states $|\Phi_{E'_{\pm}}^{1\, \alpha}\rangle$ 
defined in Eq.~(\ref{eq:chiral_u1st}). For the matrix elements of the electric field Hamiltonian one finds\cite{Trif2010}
\begin{equation}
\left|\left\langle \Phi_{E_-'}^{1\alpha} \right|  H_{d-\varepsilon}^0  \left|  
\Phi_{E_+'}^{1\alpha}  \right\rangle \right| \simeq \left|\frac{t^3}{U^3} {\mathcal{E}}e a 
\right|,
\label{eq:eEade}
\end{equation}
\begin{equation}
\left|\left\langle \Phi_{E_-'}^{1\alpha} \right|  H_{d-\varepsilon}^1  \left|  
\Phi_{E_+'}^{1\alpha}  \right\rangle \right| \simeq  \left|\frac{4t}{U}{\mathcal{E}}d_{EE} 
\right|\;.
\label{eq:EdEEde}
\end{equation}
It follows that the electric-dipole moment $d$ of the spin electric coupling is given by a combination of 
$\left|\frac{t^3}{U^3} e a \right|$ and $\left|\frac{4t}{U}d_{EE}\right|$. 

4. In the presence of an electric field, the degenerate GS chiral manifold 
$\{|\Phi_{E'_{\pm}}^{1\, \alpha}\rangle \}$ 
is replaced by the coherent linear superpositions
\begin{equation}
\left | \chi_{\pm} ^{\alpha} (\pmb{\varepsilon})\right \rangle = 
=\frac{1}{\sqrt{2}}\left(|\Phi_{E'_{+}}^{1\, \alpha}\rangle + \pm\frac{|{\bf
d\cdot \pmb{\varepsilon}|}} {\bf
d\cdot \pmb{\varepsilon}} |\Phi_{E'_{-}}^{1\, \alpha}\rangle \right)
\label{mixed_chirals}
\end{equation}
with energies 
\begin{equation}
{\rm E}_{\pm}({{\varepsilon}}) = {\rm E}_{\pm}(0) 
\pm  d\,{{{\varepsilon}}}/ \sqrt{2}
\end{equation}
Note that spin degeneracy is preserved, even when SOI is included.
The electric-field-induced splitting of the chiral GS, 
$\Delta {\rm E}({{\varepsilon}}) \equiv  
{\rm E}_{+}({{{\varepsilon}}}) - {\rm E}_{-}({{\varepsilon}})$, 
is proportional to ${{\varepsilon}}$, at least in this approximation, 
in agreement
with the effective spin Hamiltonian approach.
We will refer to the  states 
$\left | \chi_{\pm} ^{\alpha} (\pmb{{\varepsilon}})\right \rangle$ 
as {\it mixed chiral states}.
They will play a crucial role in transport.

\noindent
5. Eqs.~(\ref{eq:Hed1}) and (\ref{eq:Hed0as}) show that an electric field, in fact, 
can couple $\{|\Phi_{E'_{\pm}}^{1\, \alpha}\rangle \}$ with $|\Phi_{A'_{2}}^{1\, \alpha}\rangle$.
However this coupling, which in principle could affect Eq.~(\ref{mixed_chirals}) is not important, since these states are 
separated by
an energy of order $J$. We will therefore disregard it.

\begin{figure}[tbp]
\includegraphics[scale=0.3]{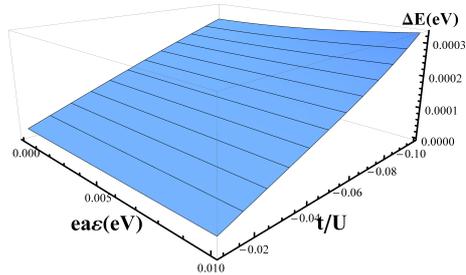}
        \caption{(Color online) Electric-field--induced splitting $\Delta {\rm E}({{\varepsilon}})$ 
                of the chiral ground state energy for a triangular molecular magnet at half-filling $(N= 3)$,  
                as a function of field strength ${\varepsilon}$
                and $t/U$. At these small/moderate values of the field, $\Delta {\rm E}({{\varepsilon}})$ 
                depends linearly on
                 $\varepsilon$.}
\label{fig:Delta_E_vs_e_t/U}
\end{figure}

\begin{figure}[tbp]
\includegraphics[scale=0.3]{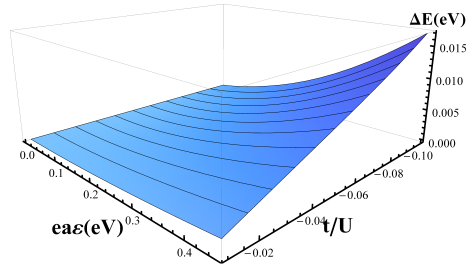}
        \caption{(Color online) The same as in Fig.~\ref{fig:Delta_E_vs_e_t/U}, but for larger values of the electric field,
                 showing a quadratic dependence of $\Delta {\rm E}({{\varepsilon}})$ due to an induced  electric dipole 
                 moment.}
\label{fig:Delta_stE_vs_e_t/U}
\end{figure}

In Figs.~\ref{fig:Delta_E_vs_e_t/U} and \ref{fig:Delta_stE_vs_e_t/U} 
we plot the computed energy splitting of the chiral GS, $\Delta {\rm E}({{\varepsilon}})$, induced
by an electric field of strength $\varepsilon$, as a function of $\varepsilon$ and $t/U$.
The splitting is, as expected, linear in $\varepsilon$ at small
fields. This is the landmark of the spin-electric coupling. However, at larger field, we find also a quadratic
dependence. It seems that, despite the large value of $U$, the system has a sizable polarizability, leading to
an rather strong induced electric dipole moment in the presence of a field. This is responsible for the
quadratic contribution in $\Delta {\rm E}({{\varepsilon}})$. 

All the calculations on the model presented in the next section are obtained by exact diagonalization
of the Hubbard model for $N= 2, 3, 4$ filling or {\it charge states}. 
It turns out, however, that for the values of the parameters relevant
for $\{ Cu_3 \}$, the perturbative results in $t/U$ are typically quite close to the exact results.

%
%

%
%
\section{Transport Model and Master Equation Approach}
\label{sec:TransportModelandMethods}

\subsection{Transport setup}

We are interested in studying quantum transport through a triangular molecular magnet (MM),
weakly coupled to conducting leads, gated, and with the possibility of an 
extra external electric field for control of the spin-electric
coupling. The transport regime that we have in mind is predominately controlled by Coulomb blockade physics. 
Later in this section we will also comment on the possibility of
employing inelastic electric tunneling spectroscopy without the presence
of charging effects.

\begin{figure}[tbp]
        \centering
                \includegraphics[scale=0.40]{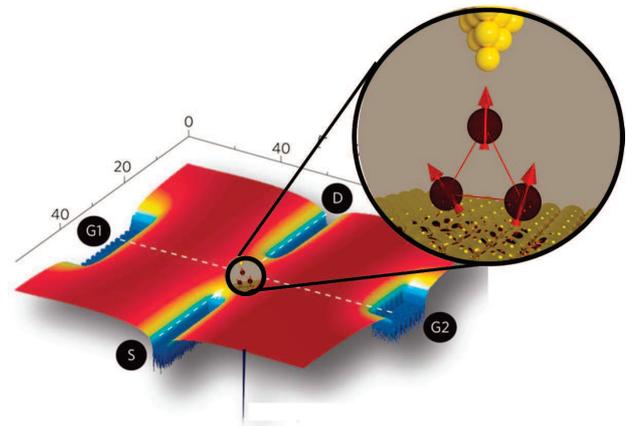}
        \caption{Schematic representation of the
transport geometry with a triangular molecular magnet. 
Picture modified from the original work by Fuechsle {\it et.al.}. Reprinted by 
permission from Macmillan Publishers Ltd: Nature Nanotechnology 7, 242–246, copyright (2012) }
        \label{fig:SZT-SET}
\end{figure}

A possible transport geometry is schematically shown in Fig.~\ref{fig:SZT-SET}.
The MM is placed on a surface (semiconducting or insulating.). Two
conducting coplanar leads 
acting as source (S) and drain (D) are constructed on
the surface,
for example using techniques recently to realize 
a single-atom transistor.~\cite{Fuechsle2012}
The molecule is weakly coupled to S and D leads via ligands.
Two in-plane gates (G1 and G2) are also 
patterned on either side of the transport channel.
The orientation
of the MM on the surface is such that the electric field from the gate is orthogonal
to the plane of the MM, and it is simply used as a capacitative coupling
to control the 
chemical potential of MM. 
Alternatively, S and D  nanoleads and gate electrodes can be constructed 
by nano-lithography by depositing metal atoms (e.g., Au) on an
insulating surface.
Finally, a STM tip is positioned in the vicinity
of the MM (see the blown-up region of the device close to the MM). This electrode is supposed to
provide another {\it strong} and {\it localized} electric field 
to manipulate the MM states via the
spin-electric coupling discussed in the previous section. 

The construction of the device described here
is 
very challenging. But we rely on recent progress in STM nano-lithography, and
especially in funcionalizing MMs on surfaces.

A second possibility is to study transport in a single-electron transistor
(SET) built in more traditional
molecular electronic device. MMs are presently being successfully
investigated with this techniques.\cite{Heersche2006,Jo2006,van_der_zant_2010,alexander10, vzant12}
Here the challenge is to provide an
independent extra gate electrode (besides the ordinary back gate)
to reliably generate an in-plane
electric field triggering the spin-electric coupling.

In the following we will assume that the following three features 
are present in our system:
(i) source and drain leads weakly coupled to the molecule, providing a bias
voltage $V_b$ for electric transport; (ii) a gate voltage generating 
a variable potential $V_g$ on the molecule able to manipulate its
charge state; (iii) a third independent 
local electric
field $\pmb{\varepsilon}$, of strengths typically attainable 
in the vicinity of a STM, with
a component in the plane of the MM.

\subsection{Hamiltonian of the transport device}

The Hamiltonian of the system, schematically represented 
in Fig.~\ref{fig:SZT-SET}, is
the sum of three terms
\begin{equation}
{\cal H}=
 {\cal H}_{L/R}+{\cal H}_{\text{mol}}+{\cal H}_{L/R}^{T}\;, 
 \label{eq:Hamiltonian}
\end{equation}
where 
\begin{equation}
{\cal H}_{L/R}=\sum_{k\alpha}\varepsilon_k^{L/R} a_{L/Rk\alpha}^{\dagger} 
a_{L/Rk\alpha}
 \label{eq:reservoirshamiltonian}
\end{equation}
describes free ({\it i.e.}, noninteracting)
electrons in the left/right conducting lead (source/drain). Here, the operator 
$a_{L/Rk\alpha}^{\dagger}$ ($a_{L/Rk\alpha}$) creates (destroys) one electron 
with wave vector $k$ and spin $\alpha$ in the left/right lead, respectively 
with energy $\varepsilon_k^{L/R}$. The tunnel junctions representing
the coupling between leads and MM are described by the 
tunneling Hamiltonian 
\begin{equation}
{\cal H}_{L/R}^{T}=\sum_{km\alpha}\left( T_{km\alpha}^{L/R} 
a_{L/Rk\alpha}^{\dagger} c_{m\alpha} + \text{H.c.} \right),
 \label{eq:tunnelinghamiltonian}
\end{equation}
where $T_{km\alpha}^{L/R}$ is the tunneling amplitude, $c_{m\alpha}^{\dagger}$ 
($c_{m\alpha}$) creates (destroys) an electron in a single particle state with 
quantum numbers $m$ and $\alpha$ inside the MM. The tunneling Hamiltonian 
${\cal H}_{L/R}^{T}$ is treated as a perturbation to ${\cal H}_{\text{mol}}$ 
and  ${\cal H}_{L/R}$.

The general form of the MM Hamiltonian is given by 
%
\begin{eqnarray}
{\cal H}_{\text{mol}}&= {\cal H}_{0}+{\cal H}_{U}+{\cal H}_{t}+{\cal H}_{\rm 
SOI}+{\cal H}_{\rm EF},
 \label{eq:dothamiltonian}
\end{eqnarray}
where 
\begin{equation}
{\cal H}_{0}=\sum_j\sum_{\alpha} \left( \epsilon_j - e\,V_g   
\right)c^{\dagger}_{j\alpha}c_{j\alpha},
\label{gateV}
\end{equation}
with $V_g$ the gate voltage. ${\cal H}_{U}= U \sum_j 
n_{j\uparrow}n_{j\downarrow}$ with $U$ the on-site Coulomb repulsion parameter 
and $n_{j\alpha}=c^{\dagger}_{j\alpha}c_{j\alpha}$ the number operator. ${\cal 
H}_{t}=  t \sum_j \sum_{\alpha} c^{\dagger}_{j\alpha}c_{j+1\alpha}+\text{H.c.}$ 
the hopping Hamiltonian with $t$ the hopping parameter.  ${\cal H}_{\rm 
EF}={\cal H}_{d-\varepsilon}^1+{\cal H}_{d-\varepsilon}^0$ the electric field 
Hamiltonian defined in Eqs. (\ref{eq:Hed1}) and (\ref{eq:Hed0as}) and ${\cal 
H}_{\rm SOI}$ the spin-orbit Hamiltonian defined in Eq. (\ref{eq:soias}).

We assume the Coulomb interaction between 
electrons in the MM and those in the environment, to be determined by a single 
and constant capacitance $C=C_L+C_R+C_g$, where $C_{L/R}$ and $C_g$ are the 
capacitances of the right/left lead and the gate electrode, respectively. 
Another assumption is that  the single-particle spectrum is independent of 
these interactions. 

Quantum transport, e.g. the calculation of the
tunneling conductance as a function of
bias and gate voltages, can now be studied by means of a quantum master
equation. General derivations of these equations have recently appeared in the
literature,\cite{Elste2005, Leijnse2008, Koller2010} together with several 
approximate solutions applied to 
SETs with quantum dots\cite{Weinmann1995} 
and molecules,\cite{Leijnse2008, Leijnse2009, Koller2010} 
including 
MMs.\cite{Elste2005,timm2006prb, elste2006prb, elste2007prb, timm2007prb}
The simplest strategy is to solve these
equations perturbatively in the tunneling Hamiltonian.\cite{Bruus2004} 
%
%

\subsection{Coulomb blockade Regime, Sequential Tunneling}
\label{sec:CoulombBlockadeRegimeSequentialTunneling}
In the regime of weak coupling between leads and molecule, transport occurs
via the so-called sequential tunneling.\cite{Bruus2004} We review here the main
characteristics of this regime an the steps leading to the calculation
of the current.\cite{Bruus2004}
In this regime the conductance of the tunnel junctions 
should be much smaller than the quantum of conductance ${\cal G}_Q=2e^2/h$.
The electron tunneling rates $\Gamma$ 
should be much smaller than the charging energy $E_c$ of the molecule and the
the temperature: 
$\hbar \Gamma \ll k_{\rm B} T \ll E_c$. 
The time between two tunneling events  $\Delta t$ is 
the longest time scale in the  regime. In particular 
$\Delta t \gg \tau_{\phi}$, where  
$\tau_{\phi}$ is the electron phase coherence. This guarantees that once the 
electron tunnels in, it has the time to loose its phase coherence 
before it tunnels out. Therefore the charge state can be treated classically 
and superposition of different charge states is not allowed. Only one-electron 
transitions between leads and molecule occur in the system. 
These transitions are characterized
by rates $\Gamma_{\ij}$, where $i,j$ are the 
initial and final system states of the system involved 
in the electron transfer. The system 
is described by stationary non-equilibrium populations $\mathcal{P}_i$ of the 
state $i$.  These occupation probabilities can be obtained from the master 
equation 
\begin{equation}
\frac{d}{dt}\mathcal{P}_i=\sum_{j(j\neq i)} \left( \Gamma_{ij} \mathcal{P}_j -  
\Gamma_{ji} \mathcal{P}_i\right) \;.
\label{eq:master}
\end{equation}
The first RHS term represents events where the electron tunnels into the 
state $i$ from the state $j$, while the second RHS term represents events where 
the electron tunnels out from the state $i$ into the state $j$. 
These probabilities obey the normalization condition
\begin{equation}
\sum_{i} \mathcal{P}_i =1   \;.
\label{eq:norma}
\end{equation}

In the steady state, the probabilities are  time-independent $d\mathcal{P}_i/dt=0$. 
Therefore, Eq.~(\ref{eq:master}) can be written as 
\begin{equation}
0=\frac{d}{dt}\mathcal{P}_i=\sum_{j(j\neq i)} \left( \Gamma_{ij} \mathcal{P}_j 
-  \Gamma_{ji} \mathcal{P}_i\right)\;.
\label{eq:zeromaster}
\end{equation}

In the regime of sequential tunneling the transition amplitudes
are computed by first-order perturbation theory in 
the tunneling Hamiltonian ${\cal H}^{T}$, 
Eq.~(\ref{eq:tunnelinghamiltonian}). Therefore the transition rates
from state $i$ to state $j$, through the left/right lead,
are given by Fermi's 
golden Rule 
\begin{equation}
\Gamma_{i \to j}^{L/R}=\frac{2\pi}{\hbar} \sum_{i,j}\left| \left\langle j  
\left| H_{L/R}^T \right|  i  \right\rangle \right|^2 W_{i}  \delta(E_j-E_i)  \;,
\label{eq:golden}
\end{equation}
where $W_i$ is a thermal distribution function and $E_j-E_i$ gives the energy 
conservation. The states $\left| i \right\rangle$ and $\left| j \right\rangle$ 
are the unperturbed system states and are defined as a product of the molecule 
and lead states $\left| i \right\rangle=\left| i_{mol} \right\rangle\otimes 
\left| i_l \right\rangle \otimes\left| i_r \right\rangle$. Transition rates 
depend on whether an electron is leaving or entering the molecule through 
the left or right lead. Inserting the tunneling Hamiltonian
Eq.~(\ref{eq:tunnelinghamiltonian}) into the Fermi's golden Rule, 
Eq.~(\ref{eq:golden}), the transition rates become\cite{Tews2004, 
Bruus2004}
\begin{equation}
\Gamma^{L/R,-}_{i  \to j}= \gamma^{L/R,-}_{ji} \left[ 1- f_{L/R}(E) \right]  \;,
\label{eq:rateminus}
\end{equation}
\begin{equation}
\Gamma^{L/R,+}_{i  \to j}= \gamma^{L/R,+}_{ji} \left[f_{L/R}(E) \right]  \;,
\label{eq:rateplus}
\end{equation}
where 
\begin{eqnarray}
\gamma^{L/R,-}_{ji}&=& \Gamma^{L/R} \sum_{m,\alpha} \left|  \left \langle j   
\left|  
c_{m,\alpha} \right|  i  \right\rangle \right|^2  
\label{eq:Sm}
\end{eqnarray}
and  
\begin{eqnarray}
\gamma^{L/R,+}_{ji}&=&\Gamma^{L/R} \sum_{m,\alpha}  \left|  \left\langle j  
\left|  \  c^{\dagger}_{m,\alpha} \right|  i  \right\rangle \right|^2 
\label{eq:Sp} 
\end{eqnarray}
are the transition matrix elements between  the states $j$ and $i$ of
the molecule (we have now dropped the label "mol"); $E=E_j-E_i$ 
is the energy difference between  molecule many-electron states, 
and $f_{L/R}(E)=\left[ 
e^{(E-\mu_{L/R})/k_BT} +1 \right]^{-1}$ is the Fermi function. Here the 
combination between the tunneling amplitudes $T^{L/R}_{m,\alpha}$  and the   
left/right lead density of states $D_{L/R}(i_{L/R})$ is assumed to be
 constant:  
$\Gamma^{L/R}=(2\pi/\hbar)\left| T^{L/R}_{m,\alpha}\right|^2 D_{L/R}(i_{L/R}) = 
(2\pi/\hbar)\left| T^{L/R}\right|^2 D_{L/R}(i_{L/R})$. The full transition 
matrix in the master equation, Eq.~(\ref{eq:master}) is the sum of all 
contributions of electrons tunneling out or into the molecule, Eqs. 
(\ref{eq:rateminus}) and (\ref{eq:rateplus}): 
\begin{equation}
\Gamma_{ij}=\Gamma^{L,+}_{ij} +\Gamma^{R,+}_{ij} +\Gamma^{L,-}_{ij} 
+\Gamma^{R,-}_{ij}  \;.
\label{eq:totalmatrixelements}
\end{equation}

The stationary rate equation, Eq. (\ref{eq:zeromaster}), is a system of linear 
equations and has to be solved numerically for a system of $n$ many-electron 
states that are taking into account. We can rewrite it as a matrix equation 
\begin{equation}
0=\sum_{j}^n \Lambda_{ij} \mathcal{P}_j,
\label{eq:probamatrix}
\end{equation}
where 
\begin{equation}
 \Lambda_{ij} = \Gamma_{ij}-\delta_{ij} \sum_{k=1}^n \Gamma_{kj}.
\end{equation}

There must exist a physical solution to Eq. (\ref{eq:probamatrix}). Therefore 
we replace the first line of of this equation by the normalization condition, 
Eq. (\ref{eq:norma}), fixing $\Lambda_{1j}=1$. Thus we can write 
\begin{equation}
\delta_{1i}=\sum_{j}^n \Lambda_{ij} \mathcal{P}_j
\label{eq:probamatrix2}
\end{equation}
instead Eq. (\ref{eq:probamatrix}). Because Coulomb blockade 
is typically studied at low temperatures
some transitions rates might become exponentially small. This leads to 
numerical problems in solving Eq.~(\ref{eq:probamatrix2}).
Then 
some of the states do not contribute and one has to develop  a convenient
truncation 
method.\cite{elste2006prb}

Finally, the current flowing through left lead coming into the molecule must be 
equal to the current flowing through right lead coming out from the molecule. 
Knowing the occupation probabilities, Eq.~ (\ref{eq:zeromaster}), the current 
through the system is defined as \cite{Weinmann1995}

\begin{equation}
I\equiv I^{L/R}=(-/+)e \sum_{i,j(j\neq i)} \mathcal{P}_j  \left( 
\Gamma^{L/R,-}_{ij} - \Gamma^{L/R,+}_{ij} \right)
\label{eq:current}
\end{equation}

This expression contains implicitly the bias and gate voltages.
Therefore IV curves can be obtained for finite values of these voltages. 
The bias derivative of the current
gives the differential conductance $G$.
When plotted as a function of the bias $V_b$, 
the current has steps in correspondence
of values of $V_b$ at which new transitions involving two contiguous
charge states are energetically allowed. At low voltages -- smaller than
the charging energy -- this is not possible and the current is blocked.
In correspondence of these transitions, the conductance as function of
$V_b$ displays peaks. When plotted simultaneously as a function 
of both  $V_b$ and  $V_g$,
the conductance displays a characteristic diamond pattern, the so-called
stability diagram: inside
each diamond a given charge state is stable and the current is blocked.

%
%

\subsection{Cotunneling Regime}
\label{sec:Cotunneling Regime}

When the coupling to the leads becomes stronger the description
of transport based on incoherent sequential tunneling is no longer
enough. In particular higher-order tunneling processes in which the electron 
tunnels coherently through classically forbidden charge states.
As a result, for values of the voltages where sequential tunneling predicts
a blocking of the current, a small leakage current is in fact possible
though these processes.\cite{Bruus2004}
The simplest example of these processes is second order in
the tunneling Hamiltonian, and it is known as cooperative tunneling
or cotunneling. Typically for the cotunneling regime 
$ k_{\rm B} T < \hbar \Gamma \ll  E_c$.

Cotunneling can be either elastic or inelastic. In the former  case
the energies of the initial and final state are the same, while in the latter 
the energies are different.  Signatures for these processes have also been 
observed in 
single-molecule junctions.\cite{Heersche2006,Jo2006,van_der_zant_2010} 
Beyond the sequential tunneling regime, 
the tunneling Hamiltonian must be replaced 
by the $T$-matrix, which is given by\cite{Bruus2004} 
\begin{equation} 
T= \mathcal{H}^T+ \mathcal{H}^T \frac{1}{E_j-\mathcal{H}_0+i\eta}T\;,
\end{equation}
where $E_j$ is the energy of the initial state $\left| j \right\rangle\left| n 
\right\rangle$, where $\left| j \right\rangle$ refers to the equilibrium state 
on the left and right lead and $\left| n \right\rangle$ is the initial 
molecular state, $\eta=0+$ is a positive infinitesimal and $\mathcal{H}_0= 
\mathcal{H}_{mol}+\mathcal{H}_{L/R}$.   To second order, the transition rates 
from state $\ket{j}\ket{n}$ to  $\ket{j'}\ket{n'}$ with an electron tunneling 
from lead $\alpha$ to the lead $\alpha'$ are given by 
\begin{eqnarray}
\Gamma_{\alpha\alpha'}^{nj;n'j'}&=&\frac{2\pi}{\hbar}\left|  \bra{j'}\bra{n'}  
\mathcal{H}^T \frac{1}{E_{jn}-\mathcal{H}_0+i\eta} \mathcal{H}^T \ket{n}\ket{j} 
\right|^2 \nonumber \\
&&\times \delta(E_{j'n'}-E_{jn}) \;,
\label{eq:ratecot}
\end{eqnarray}
where $E_{j'n'}$ and $E_{jn}$ are the energies of the final and initial states, 
respectively. Here  $\ket{j'}\ket{n'}= a_{\alpha' \mathbf{k}'  
\sigma'}^{\dagger} a_{\alpha \mathbf{k}  \sigma}\ket{j}\ket{n'}$. 
Inserting the tunneling Hamiltonian, Eq.~(\ref{eq:tunnelinghamiltonian}),  in 
last equation and after some algebra (see Appendix \ref{sec:appendixgammasum}) 
one can get the expression for the transition rates for processes from lead 
$\alpha$ till lead $\alpha'$ and from molecular state $\left| n \right\rangle$ 
to the state $\left| n' \right\rangle$:
\begin{widetext}
\begin{eqnarray}
\Gamma_{\alpha\alpha'}^{n;n'} &=&
\sum_{\sigma\sigma'} \gamma_{\alpha}^{\sigma}\gamma_{\alpha'}^{\sigma'}
\int d \varepsilon 
 f \left( \varepsilon -\mu_{\alpha} \right)
\left(1-f\left(\varepsilon +\varepsilon_n -\varepsilon_{n'} -\mu_{\alpha'}
\right)\right)
   \nonumber 
  \\
 &&\ \ \ \ \ \  \ \ \ \ \ \ \ \ \   \ \ \ \ \ \ \ \ \   \ \ \ \ \ \ \ \ \ \times
\left|
\sum_{n''} 
\left\{
\frac{ A_{n''n'}^{\sigma *} A_{n''n}^{\sigma'}}
{\varepsilon-\varepsilon_{n'}+\varepsilon_{n''}+i\eta}  +
\frac{ A_{n'n''}^{\sigma'}A_{nn''}^{\sigma *}}
{\varepsilon+\varepsilon_{n}-\varepsilon_{n''}+i\eta}
\right\} \right|^2,  
\label{eq:gammasum}
\end{eqnarray}
\end{widetext}
where $\sigma$ is the electron spin, $f(\varepsilon)$ is the Fermi distribution 
function, $\mu_{\alpha}$ is the chemical potential of the lead $\alpha$, 
$\mu_L-\mu_R=-eV/2$, $\left| n'' \right\rangle$ is a virtual state, 
$A_{ij}^{\sigma'} =\bra{i} c_{\sigma'} \ket{j} $ and 
$A_{ij}^{\sigma*}=\bra{j}c_{\sigma}^{\dagger} \ket{i}$. Here 
$\gamma_{\alpha}^{\sigma}$ is the tunneling amplitude. Note that 
$|n\rangle$ and $|n'\rangle$ are 
states with the same number of particles. We have not taken into account 
processes changing the electron number 
by $\pm$2 units.\cite{Koch2006,Leijnse2009}  

The transition rates in Eq.~(\ref{eq:gammasum}) cannot 
be evaluated directly because of the second-order 
poles in the energy denominators. A regularization scheme has been carried 
out to fix these divergences  and obtain the cotunneling 
rates.\cite{Turek2002,Koch2004}  Here it is important to mention that these 
divergences are, in fact, an artifact of the 
$T$-matrix approach 
rather than a real physical problem. The fourth-order Bloch-Redfield quantum 
master equation (BR) and the real-time diagrammatic technique (RT) approaches 
to quantum transport have been developed to avoid any divergences and therefore 
no {\it ad hoc} regularization to cotunneling is 
required.\cite{Leijnse2008,Koller2010} 
Nevertheless, the $T$-matrix approach agrees 
with these two approaches and gives good reasonable results deep inside the 
Coulomb blockade region.\cite{elste2007prb} 
We expect to catch all the relevant physics for our 
system with the $T$-matrix approach. After the regularization scheme is 
implemented, we get the tunneling rates defined as (see Appendix 
\ref{sec:rightrates})
\begin{widetext}
\begin{eqnarray}
\Gamma_{\alpha\alpha'}^{n;n'}
&=& 
\sum_{\sigma\sigma'} \gamma_{\alpha}^{\sigma}\gamma_{\alpha'}^{\sigma'}
 \psqrt{
\sum_k  \leri{
 A^2 J(E_1,E_2,\ve_{ak})
+B^2 J(E_1,E_2,\ve_{bk} )  }
\right. +2
\sum_q\sum_{k\neq q}  
   A_k A_{q} I(E_1,E_2,\ve_{ak},\ve_{aq}) 
\nonumber \\
&&+2
\sum_q\sum_{k\neq q}    B_{k}B_{q} I(E_1,E_2, \ve_{bk},\ve_{bq})
+\left.	
2\sum_q\sum_{k}
A_k B_{q} I(E_1,E_2,\ve_{ak},\ve_{bq}) }
\label{eq:cotunnelingrates}  
\end{eqnarray}
\end{widetext}
where $A_k=A_{kn'}^{\sigma *} A_{kn}^{\sigma'}$, 
$B_k=A_{n'k}^{\sigma'}A_{nk}^{\sigma *}$, $\ve_{ak}=\ve_{n'}-\ve_{k}$, 
$\ve_{bk}=\ve_{k}-\ve_{n}$, $E_1=\mu_{\alpha}$ and 
$E_2=\mu_{\alpha'}+\ve_{n'}-\ve_{n}$. Here $I$ and $J$ are integrals that come 
out from the regularization scheme, and are defined in Eqs.~(\ref{eq:I}) and 
(\ref{eq:J}), respectively.

The complete master equation, including both sequential and cotunneling 
contributions, finally reads
\begin{eqnarray}
\frac{d}{dt}\mathcal{P}_i 
&=&
 \sum_{j(j\neq i)} \left( \Gamma_{ij} \mathcal{P}_j -  \Gamma_{ji} 
\mathcal{P}_i\right) \nonumber \\
&&  \ \ \ \ \ \ \ \ \ \ +\sum_{\alpha \alpha' j}  \left( \Gamma_{\alpha 
\alpha'}^{ji} \mathcal{P}_j -  \Gamma_{\alpha \alpha'}^{ij} \mathcal{P}_i 
\right)
 \;,
\label{eq:mastercot}
\end{eqnarray}
and the current through the system is now given by
\begin{eqnarray}
I&\equiv& I^{L/R}=(-/+)e \sum_{i,j(j\neq i)} \mathcal{P}_j  \left( 
\Gamma^{L/R,-}_{ij} - \Gamma^{L/R,+}_{ij} \right) \nonumber \\
&&+(-/+)e \sum_{i,j(j\neq i)} \mathcal{P}_j	  \left( \Gamma^{ji}_{LR/RL} - 
\Gamma^{ij}_{RL/LR} \right)
\label{eq:currentcot}
\end{eqnarray}

As mentioned above, cotunneling gives rise to a small current
inside a Coulomb-blockade diamond region 
of a given charge state. At small values of the
bias voltage, smaller than any excitation energies for the given
charge state, we are in the regime of {\it elastic cotunneling} and
the current is proportional to the bias voltage. At voltages 
corresponding to the transition energy to the first excited state
of the {\it same} charge state, 
a new cotunneling transport channel becomes available and the slope
of the linear dependency of the current increases. This signals the
first occurrence of {\it inelastic cotunneling}. Upon further
increasing the bias, other upward changes of the slope of the current
occur in correspondence to energies at which higher excited states become
available. 
It follows
that the differential conductance displays {\it steps}  
that resemble the IV curve in the sequential tunneling regime. 
Note however, that the nature of the two curves is very different:
at low bias the conductance is finite (elastic cotunneling).
Furthermore the width of the steps in the cotunneling conductance gives
the energy difference between states of the {\it same} charge
state, fixed by the specific Coulomb diamond of the 
stability diagram. Therefore, cotunneling is an excellent tool to 
investigate {\it directly} the excitation energies of a given charge state.
Indeed cotunneling spectroscopy has been used 
to investigate electronic, vibrational and magnetic excitations
in nanostructures such as a-few-electron semiconductor quantum 
dots,~\cite{Schleser2005} carbon nanotube quantum dots,~\cite{Holm2008, 
Paaske2006} metallic carbon nanotubes,~\cite{Sapmaz2005} and single-molecule 
junctions.~\cite{Roch2008,Parks2007,Osorio2007} 

At this point, before analyzing the transport results of our model,
it is useful to make a connection with inelastic electron tunneling
spectroscopy (IETS), studied for example by electron tunneling from a scanning tunneling spectroscope (STM) tip
through a molecule adsorbed on a surface\cite{Galperin2007, Reed2008}. 
The reader familiar with IETS
easily recognizes that the differential conductance versus applied voltage
for this case is very similar to the cotunneling conductance of
Coulomb blockade. This similarity is not accidental: the physics is
essentially the same in both cases, since it involves the coherent electron
tunneling through a finite system, whose internal degrees of freedom
(e.g., vibrational, magnetic and electronic) can be excited by the
process. The mathematical formulation of the problem is very similar
in the two cases. There is one noticeable difference. In IETS by STM
the coupling between the molecule and the (conducting) substrate
is much stronger that the coupling between the STM tip and the molecule.
Therefore typical IETS setups can be viewed as strongly asymmetric
Coulomb-blockade systems, when these are studied in the cotunneling
regime.

These considerations suggest an alternative way to investigate
the spin-electric coupling in triangular MMs via quantum transport.
In the setup of Fig.~\ref{fig:SZT-SET} we can imagine that transport
through the MM occurs between the STM  and the substrate.
on which the MM is placed.
Now the gates and leads constructed on the surface could provide
the external electric field responsible for the spin-electric tunneling.
For this purpose the plane of the triangular MM should be parallel to the
surface of the substrate.
In this case the detection and coherent manipulation of the low-energy
chiral states of the MM would occur by means of IETS.

%
%

%
%
%
%
%
%
%
%
%
%
%
%
%
%
%
%
%
%
%
%

%
\section{Results and discussion}
\label{sec:Results}

We now discuss quantum transport for the setup 
of Fig.~\ref{fig:SZT-SET}  We first construct the relevant low-energy many-body states for the
charge states containing $N = 2, 3, 4$ electrons. For this purpose
we use he Hubbard model introduced in Sec.~\ref{sec:Spinelectriccoupling}.
The parameters of the model are taken from the first-principles studies
on the $\{ Cu_3 \}$ triangular molecular magnet (MM) by Ref.~\onlinecite{Nossa2012}.
We have $t=-51$ meV, $U=9.06$ eV, $\lambda_{\rm SOI}=0.4$ meV.
The model is solved exactly for $N = 2, 3, 4$.
We label the many-body states with their electron number $N$ (the charge state), 
total spin $S$ and $z$-component of the total 
spin $S_z$\footnote{In principle, because of the presence of the spin-orbit 
interaction, states with different total $S$ are coupled. However the mixing
is of the order of the Dzyaloshinskii-Moriya (DM) parameter $D \propto t \lambda_{\rm SOI}/U$,
which, for the parameters used here, is very small on the scale of the
exchange constant separating states with different $S$. Therefore, in practice,
$S$ and $S_z$ are good quantum numbers.} In case of additional degeneracy,
we will use additional quantum numbers to specify the states, e.g., for the the chiral degeneracy for the $N=3$ 
ground state (GS), we will 
add $E'_{\pm}$.

%
%
%
\begin{figure}[tbp]
	\centering
		\includegraphics[scale=0.40]{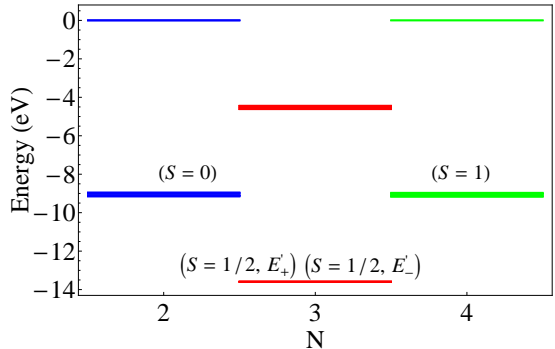}
	\caption{(Color online) Low-energy spectrum  
of the triangular 
molecular magnet described by the Hubbard model, 
Eq.~(\ref{eq:hubbard_soi}), for different charge states or electron filling,
$N= 2, 3, 4$. Here the Hubbard model parameters,
$t=-0.051$, $U=9.06$, $\lambda_{\rm SOI}=0.0004$ (all in eV),
are taken from first-principles calculations\cite{Nossa2012} 
for the $\{ Cu_3 \}$ molecular magnet. A gate voltage $V_g=U/2$ has been added 
to rigidly shift the
spectrum of the system for a given $N$. 
The total spin of the ground state (GS) for the 
different charge states
is indicated in parenthesis. The GS 
for the $N=3$-particle system corresponds to the chiral states, $E'_{\pm}$,
defined in Eq.~(\ref{eq:chiral_u1st}).}
	\label{fig:E234}
\end{figure}

The low-energy levels for the three contiguous charge states are shown in
Fig.~\ref{fig:E234}. 
To the energies calculated with the Hubbard model, we have added
a gate voltage term $-eV_g\, N = -U/2\, N$, which shifts rigidly the spectra
of the different charge states with respect to each other. 
This choice makes the spectra of the $N= 2$ and $N = 4$ charge states
more symmetric with respect to the $N= 3$ states. We will also use this value
of the gate voltage below, in the study of cotunneling transport, to  
make sure that the system is stable in the middle of the $N=3$ Coulomb diamond.

For the present choice of the Hubbard parameters, 
these states are well described
by the perturbative analysis of Sec.~\ref{sec:Spinelectriccoupling}.
As discussed there, the GS for the $N=3$ charge state 
(lowest middle line) is four-fold
degenerate, and it corresponds to the states
defined in Eq.~(\ref{eq:chiral_u1st}). 
In Fig.~\ref{fig:E234} the same line denotes the position of the 
$S= 3/2$ excited state, whose separation from the GS is not visible on this
energy scale.

\begin{figure}[tbp]
	\centering
		\includegraphics[scale=0.55]{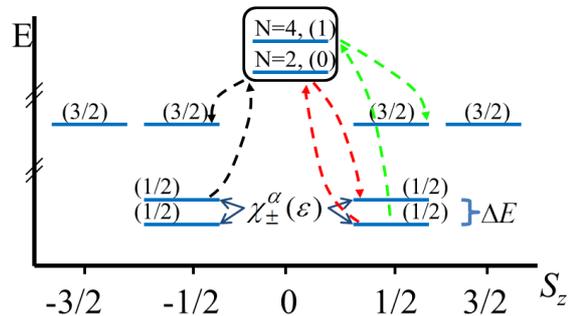}
	\caption{(Color online). Schematic energy diagram of a triangular 
molecular magnet in the presence of an external electric field 
$\varepsilon$. Only the ground state (GS) of the $N=2,3,4$-particle system 
and the lowest excited states of the
$N=3$ system are included. 
The numbers in 
parenthesis corresponds to the total spin $S$. 
The electric field lifts the $N=3$ 
GS degeneracy, and mixes the chiral states 
defined in Eq.~(\ref{eq:chiral_u1st}). 
The ``mixed chiral states'', are now labeled by $\chi^{\alpha}_{\pm}$,
with $\chi^{\alpha}_-$ being the GS.
The GS splitting $\Delta E$ is linear in $\varepsilon$ at low fields.
Here we have used the same parameters 
of Fig.~\ref{fig:E234}, plus $ea\varepsilon=0.487$eV, 
and $d_{EE}\varepsilon=0.1ea\varepsilon$. The electric field is applied
in the plane of the triangle, perpendicularly to line joining vertexes 1 and 2
of the triangle. Also shown in the figure with dashed-colored lines are
allowed inelastic cotunneling transitions,  
occurring via $N = 2, 4$ virtual ground states, $2_0$ and $4_0$, respectively. 
Red, black and green dashed lines correspond 
to transitions: $\chi_-\leftrightarrow \chi_+$ ($\Delta E$), 
$\chi_+\leftrightarrow S=3/2$ and $\chi_-\leftrightarrow S=3/2$, 
respectively.}  
	\label{fig:Ener}
\end{figure}

We now consider the presence of a strong and localized electric field, 
generated, for example,
by a scanning tunneling spectroscope (STM) tip positioned nearby the MM. We will consider values of
$\varepsilon$ up to a maximum equal 0.1V/\AA, which can be
easily attained with a STM.\cite{Mingo1998,Stokbro1998}
For a $\{ Cu_3 \}$ MM, the distance between magnetic ions is
$a=4.87$\AA. For a spin-electric coupling strength $d = ea$,
which is the maximum value estimated in Ref.~\onlinecite{Trif2008},
the energy scale $ea\varepsilon$ is equal to 0.487 eV 
when $\varepsilon = 0.1$ V/\AA.
As discussed in Sec.~\ref{sec:Spinelectriccoupling}, 
we model the effect of the electric field in the
Hubbard approach via the parameters $a$, $d_{EE}$, $d_{AE}$ entering the 
single-particle Hamiltonians in Eqs.~(\ref{eq:Hed1}) and (\ref{eq:Hed0as}).
Here we take $d_{EE}= 0.1 ea$ and $d_{AE} =0$. 
The effect of the field on the low-energy spectrum of the MM is shown
in Fig.~\ref{fig:Ener}, with the expected splitting and mixing of the 
GS chiral states for the $N=3$ charge state. In the absence of spin orbit interaction (SOI)
the ``mixed chiral states'' 
$|\chi^{\alpha}_-(\varepsilon)\rangle$ and 
$|\chi^{\alpha}_+(\varepsilon)\rangle$ 
(with $|\chi^{\alpha}_-(\varepsilon)\rangle$ being the GS)
are still spin ($\alpha = \pm 1/2$) degenerate. As we saw, their splitting  
$\Delta E(\varepsilon)$ is proportional to $\varepsilon$.
It is interesting to note that, the (small) spin-orbit coupling given
in Eq.~(\ref{eq:soias}), mixes a little bit $|
\chi^{\alpha}_-(\varepsilon)\rangle$ and
$|\chi^{\alpha}_+(\varepsilon)\rangle$. However, since the effect is the
same for $\alpha = \pm 1/2$, the double degeneracy of the GS and the first
excited state is preserved, and the splitting remains of the order of 
$\Delta E(\varepsilon)$.

Shown on the same figure are also the four-fold degenerate (N= 3, $S=3/2$) excited state
and the $N=2$ and $N=4$ GS, having spin $S=0$ and $S=1$ respectively.
The $N=2(4)$ 
GS has total spin $S=0(1)$ and spin projection $S_z=0(0)$. The rest 
of the energy spectrum is not shown in Fig.~\ref{fig:Ener}. 

\begin{figure}[tbp]
\includegraphics[scale=0.25]{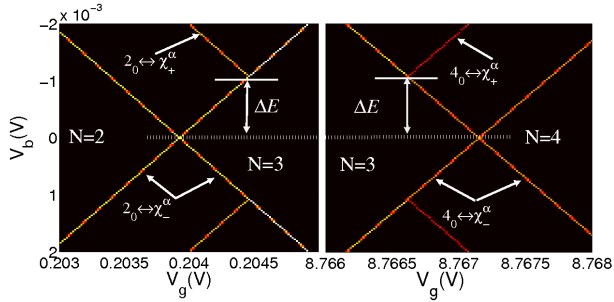}
        \caption{(Color online) Differential conductance as a function of the
bias and gate voltages in the sequential tunneling regime (stability diagram),
showing the Coulomb diamonds for three contiguous charge states $N=2, 3, 4$.
Only the corners of the diamonds are shown. The arrows indicate the electron transitions
responsible for peaks in the conductance. States are labeled following the notation of Fig.~\ref{fig:Ener}.
The calculations are done for a symmetric device 
at temperature $T \sim 10^{-2} K$ ($k_{\rm B} T \sim 0.001 {\rm meV}$).
The parameters for the Hubbard model are the same of those in Fig.~\ref{fig:Ener},
A local electric field $\varepsilon = 0.1$ V/\AA, is also included, causing a spin-electric coupling of the
$N=3$ chiral states and a ground state splitting $\Delta {\rm E}$.}
\label{fig:CB_stability_diagram}
\end{figure}

In Fig.~\ref{fig:CB_stability_diagram} we plot the Coulomb blockade stability diagram,
that is, the differential conductance in the sequential tunneling regime as a function of bias and gate voltages.
The calculations are done for a symmetric device, where the capacitances and tunneling resistances 
for the two junctions are the same. 
The temperature is taken to be  
$T \sim 10^{-2} K$ ($k_{\rm B} T \sim 0.001 {\rm meV}$).
The calculations are done for the parameters of Fig.~\ref{fig:Ener}, and an electric field $\varepsilon = 0.1$ V
/\AA\ is
included, generating a GS splitting $\Delta E $ for the $N=3$ charge state.
The picture displays familiar Coulomb diamonds for the three contiguous charge states $N = 2, 3, 4$, 
inside which the current is zero.
The lines delimiting these diamonds represent the onset of tunneling current, where the conductance has  peaks.
They correspond to real transitions between states of two contiguous charge states $N \rightarrow N\pm 1$.
The first lines where this happens involve the transition between the corresponding GSs. Other lines, parallel 
to these, 
involve
transitions between excited states, which become occupied out of equilibrium. We do not include any energy or 
spin
relaxation mechanism in these calculations.

\begin{figure}[tbp]
\includegraphics[scale=0.4]{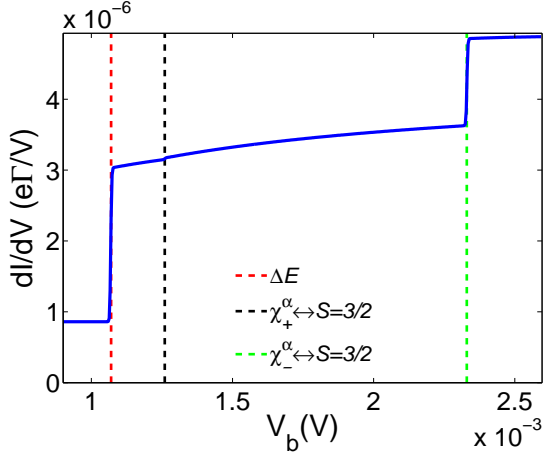}
	\caption{(Color online) Cotunneling differential conductance as a function of the 
bias voltage for parameters as in Fig.~\ref{fig:E234}. The states involved are labeled 
as in Fig. \ref{fig:Ener}. At low voltage, transport is through
elastic cotunneling. The red-dashed line corresponds 
to the first onset of inelastic cotunneling, due to the occupation of the lowest excited state 
$(N= 3, \chi_+^{\alpha})$,
through a virtual transition $(N= 3, \chi_-^{\alpha})$ ground state (GS) $\rightarrow$  $(N=2, S=0),(N=4, S=1)$ 
GSs.
The black-dashed line and green-dashed line indicate inelastic cotunneling steps caused by the final occupation 
of the 
$(N=3, S=3/2)$ excited state via the virtual transitions 
from $(N= 3, \chi_{\pm} ^{\alpha})$ to the $(N=2, S=0),(N=4, S= 1)$ GSs.}
\label{fig:dIdVcot}
\end{figure}

We now consider transport in the cotunneling regime. In Fig.~\ref{fig:dIdVcot} we plot the differential 
conductance
as a function of the bias voltage $V_b$,
for $V_g= U/2$, which locates the system in the middle of $N=3$ 
Coulomb diamond, that is, deep inside the Coulomb blockade regime.
Here the sequential tunneling current is suppressed, and transport is entirely due to cotunneling.
The conductance is nonzero even at zero bias, due to elastic cotunneling. At $V_b \approx 1.1$ meV,
the conductance has a first step, indicated by the red dashed line.
The step signals the onset of inelastic cotunneling, which takes place when the bias voltage
provides enough energy for the final occupation of the lowest excited state of the $N=3$ charge state
$(N= 3, \chi_+^{\alpha})$, via the virtual 
transition from the $(N= 3, \chi_-^{\alpha})$ GS to the $(N=2, S=0),(N=4,S=1)$ GSs. Therefore, the width 
of this first step provides a direct estimate of the energy splitting between the mixed chiral states,
$(N= 3, \chi_+^{\alpha})$ and $(N= 3, \chi_-^{\alpha})$, caused by the spin-electric coupling. 
Increasing further the bias, other two cotunneling
channels open up, causing the appearance of two other steps in the conductance. The first one, quite small,
indicated by the black dashed line, is related with the first occupation of the $(N=3, S=3/2)$ excited state,
which occurs via the virtual transition from the $(N= 3, \chi_+^{\alpha})$ excited state to the $(N=2, S=0),(N=4
, S=1)$ GSs.
Note that the state $(N= 3, \chi_+^{\alpha})$ is already occupied because of the first inelastic cotunneling
transition.
The second (higher) step, indicated by a green dashed line, is again due to the occupation of the
 $(N=3, S=3/2)$ as a final state, but though the virtual transition from the  $(N= 3, \chi_-^{\alpha})$ GS
to the $(N=2, S=0),(N=4, S=1)$ GSs.

\begin{figure}[tbp]
\includegraphics[scale=0.3]{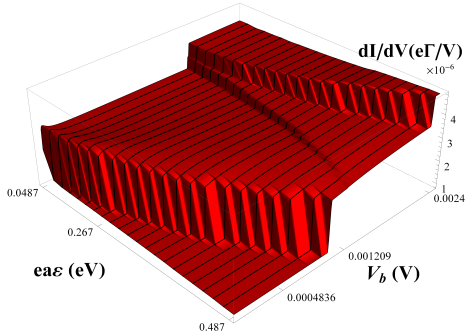}
	\caption{Cotunneling differential conductance as a function of the bias voltage and the local 
 electric field triggering the spin-electric coupling.}
\label{fig:Efields}
\end{figure}

The cotunneling conductance pattern depends on the external electric field $\varepsilon$. 
In Fig.~\ref{fig:Efields} we plot the conductance as function of the external electric field, $\varepsilon$ and 
bias voltage, $V_b$.
As expected, the value of the voltage where the first inelastic step occurs increases with
the field. Variations of the position of the other two inelastic steps in the conductance 
as a function of  $\varepsilon$ are also visible: at low fields, where the splitting of the chiral GS 
vanishes, the other two inelastic steps involving the $(N= 3, S= 3/2)$ excited state
occur at the same bias. Surprisingly, the height of the inelastic steps is {\it not} strongly
affected by the electric field. The only exception is the second  step, whose height becomes
very small at the maximum value of $\varepsilon$, as also shown in Fig.~\ref{fig:dIdVcot}.

\begin{figure}[tbp]
\includegraphics[scale=0.4]{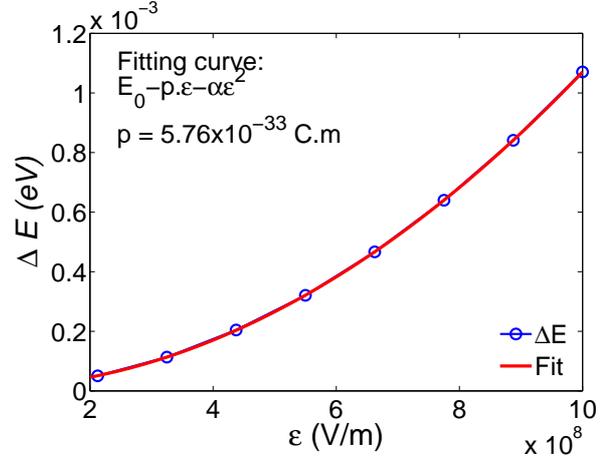}
	\caption{Energy splitting of the $N=3$ chiral ground state, $\Delta E$, caused by the spin-electric coupling, 
as a function of 
the external electric field. The values of $\Delta E$ correspond to the position of the first 
conductance step in Fig.~\ref{fig:Efields}. The fitting curve contains a linear term 
proportional to a
dipole moment $p= 5.76\; 10^{-33}$ C\; m, in agreement with the first-principles
 calculations on $\{ Cu_3 \}$ molecular magnet of Ref.~\onlinecite{Islam2010}.}  
\label{fig:fit}
\end{figure}

In Fig.~\ref{fig:fit} we plot $\Delta E$, extracted from the position of first inelastic step, as a function
of $\varepsilon$. 
A polynomial fitting
of $\Delta E$ vs. $\varepsilon$ finds, besides a quadratic contribution due to an induced electric
dipole moment, a linear term, which dominates at low fields, and it is the
landmark of the (linear) spin-electric coupling. Interestingly, the extracted 
value of the proportionality coefficient of the linear term, i.e. the ``electric dipole moment''
$p = d/\sqrt{2}$, is equal to $5.76\; 10^{-33}\; {\rm C\; m}$, which is consistent with 
the value found previously by $ab$-initio methods for $\{ Cu_3 \}$ molecular magnet.\cite{Islam2010} 
This indicates that our choice of the spin-electric parameter $d_{EE} = 0.1 ea$ 
(see Eqs.~(\ref{eq:dEEpar}) and (\ref{eq:Hed1})\ ) is
in the right ballpark.
In principle, the curve plotted in  Fig.~\ref{fig:fit} can be directly extracted from 
experimental measurements of the conductance in the cotunneling regime. From this curve,
the strength of electric dipole moment $d$ can be estimated. 

\begin{figure}[tbp]
\includegraphics[scale=0.35]{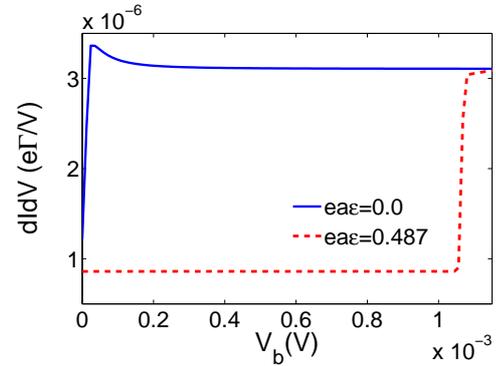}
	\caption{(Color online) Cotunneling differential conductance versus bias voltage with 
(dashed red line) and without (blue solid line) external electric field, 
causing the spin-electric coupling. 
Here we have 
used the same parameters of Fig.~\ref{fig:Ener}.}  
\label{fig:wwoutsoi}
\end{figure}

The cotunneling conductance for both $\varepsilon =0$ (blue line) and
$\varepsilon =0.1 $ V/\AA\ (red dashed line) is plotted in Fig.~\ref{fig:wwoutsoi}.
At zero field, the splitting of the $N=3$ GS, controlling the onset of inelastic cotunneling,
is brought about only by the
SOI-induced Dzyaloshinskii-Moriya interaction, which splits the chiral states
without mixing them. This splitting is predicted to be very small, both experimentally\cite{Trif2008}
($\Delta_{\rm SOI} = 0.04$ meV) and theoretically ($\Delta_{\rm SOI} = 0.02$ meV)\cite{Nossa2012}.
The value extracted from the cotunneling conductance of  Fig.~(\ref{fig:wwoutsoi}) is 
consistent with this estimate. A measurement of this splitting from cotunneling experiments is also in 
principle possible but probably very challenging.
The value of the elastic cotunneling conductance is slightly larger when the $\varepsilon$-field is absent
than in the presence of the field. However value of the inelastic conductance  is the same with and without field.
The fact that inelastic cotunneling sets in at very different thresholds with and without field suggests
the possibility of using this system as a switching device, which can be controlled electrically, possibly 
by a time-dependent field.

%
%
%
%
%
%
%
%
%
%

\section{Conclusions}
\label{sec:Conclusions}

In summary, we have carried out a theoretical study of quantum transport
through an antiferromagnetic triangular molecular magnet (MM),
in a single-electron transistor setup. The interplay of spin frustration and lack of inversion symmetry
in this MM  is responsible for the existence of an efficient spin-electric coupling, which can affect
the non-linear transport regime. When a strong localized electric field is applied to the molecule,
the spin-electric coupling causes a splitting between the two doubly-degenerate spin chiral states that compose
the ground state of the MM.  We have shown that this energy splitting and, consequently the strength
of the spin-coupling, should be directly accessible through experiments 
by measuring the inelastic cotunneling conductance in the Coulomb blockade regime.
Both single-electron transistors (SETs) used in molecular spintronics and inelastic electron tunneling
spectroscopy (IETS) of molecules on surfaces addressed with a STM could
be employed to study this effect. 

Our theoretical approach was based on a Hubbard model,\cite{Trif2008, Trif2010} where
the spin-electric coupling can be described in terms of a few microscopic parameters derivable
from first-principles calculations. We have shown that the value of the strength of spin-electric coupling
estimated from tunneling transport is consistent with the value calculated by
first-principles methods.\cite{Islam2010}

Antiferromagnetic molecules, like the one considered here, characterized by ground states composed of chiral 
pairs
of spin-1/2 doublets, could be used to create pairs of  quasi-degenerate qbits. 
The possibility of coherently coupling these two
qbits electrically and detecting their quantum superposition state in electronic transport is
an interesting topic that should further investigated. 

The effect of an external magnetic field,
not considered in this paper, can be used for gaining full control of the ground-state manifold.
Furthermore, higher excited states of the system can play a role as auxiliary states employed to perform 
quantum gates.
As we have shown in our study of the cotunneling conductance (see Fig.~\ref{fig:Efields}), 
these higher states can also
be manipulated electrically and brought closer to or further apart from the ground-state manifold.
One important issue that we have not discussed in this work is the effect of spin relaxation
on transport. This certainly plays
a crucial role in determining the robustness of the coherent superposition induced by the electric field.

%
%

\section*{Acknowledgment}
\label{sec:Acknowledgment} We would like to thank D. Loss and D. Stepanenko for several
important discussions and clarifications on the spin-electric coupling in molecular magnets, 
and M. Islam for an ongoing collaboration on the same subject. We would like to thank 
Dr. Magnus Paulsson for his help in developing the codes used in this work.
 This work was supported by the School of
Computer Science, Physics and Mathematics at Linnaeus University,
the Swedish Research Council under Grants No: 621-2007-5019 and
621-2010-3761, and the NordForsk research network 080134
``Nanospintronics: theory and simulations".

\appendix

\onecolumngrid

%
%
%

\section{\label{sec:appendixgammasum}Explicit derivation of 
Eq.~(\ref{eq:gammasum})}

Here we demonstrate the Eq. (\ref{eq:gammasum}). We study the transition rates up to 
four order. The transition rate from state $\ket{j}\ket{n}$ to  
$\ket{j'}\ket{n'}$ with one electron tunneling from lead $\alpha$ to the lead 
$\alpha'$ is given by 
\begin{equation*}
\Gamma_{\alpha\alpha'}^{nj;n'j'}=\frac{2\pi}{\hbar} \ppar{    \bra{j'}\bra{n'}  
\sH^T \fo{E_{jn}-\sH_0+i\eta} \sH^T \ket{n}\ket{j}     }^2 
\delta(E_{j'n'}-E_{jn}) \;,
\end{equation*}
where $E_{j'n'}$ and $E_{jn}$ are the energies of the final and initial states, 
respectively. $\sH_T=\sum\limits_{\alpha=L,R}t_{\alpha} \sum\limits_{\bk 
\sigma} \leri{  a_{\alpha \bk  \sigma}^{\dagger} c_{\sigma} 
+c_{\sigma}^{\dagger} a_{\alpha \bk  \sigma} }$ is the tunneling Hamiltonian 
Eq. (\ref{eq:tunnelinghamiltonian}) with $T_{km\alpha}^{L/R}=t_{\alpha}$. 
$\sH_0=\sH_{mol}+\sH_{leads}$ and $\eta$ is a positive infinitesimal number. 
Here  $\ket{j'}\ket{n'}= a_{\alpha' \bk'  \sigma'}^{\dagger} a_{\alpha \bk  
\sigma}\ket{j}\ket{n'}$. $\ket{j}$ ($\ket{n}$) refers to the equilibrium state 
of the left and right Fermi sea (molecule). The total cotunneling rates for 
transitions that involve virtual transitions between two $n,n'$-occupied 
molecule  states are then given by
\begin{eqnarray}
\Gamma_{\alpha\alpha'}^{nj;n'j'}
&=&
\frac{2\pi}{\hbar} \sum_{\bk\bk'\sigma\sigma'}
\ppar{    \bra{j}\bra{n'} a_{\alpha \bk  \sigma}^{\dagger} a_{\alpha' \bk'  
\sigma'}  
\sum_{\alpha'''}t_{\alpha'''}^* \sum_{\bk''' \sigma'''}  \leri{ a_{\alpha''' 
\bk'''  \sigma'''}^{\dagger} c_{\sigma'''} 
+c_{\sigma'''}^{\dagger} a_{\alpha''' \bk'''  \sigma'''}} 
\right. \nonumber \\
&&\times \left.
\fo{E_{jn}-\sH_0+i\eta} 
\sum_{\alpha''}t_{\alpha''} \sum_{\bk'' \sigma''} \leri{  
a_{\alpha'' \bk''  \sigma''}^{\dagger} c_{\sigma''} 
+c_{\sigma''}^{\dagger} a_{\alpha'' \bk''  \sigma''} }
 \ket{n}\ket{j}     }^2 \delta(E_{j'n'}-E_{jn}) 
 \nonumber \\
 &=&
\frac{2\pi}{\hbar}\sum_{\bk\bk'\sigma\sigma'}
\ppar{    \bra{j}\bra{n'} a_{\alpha \bk  \sigma}^{\dagger} a_{\alpha' \bk'  
\sigma'}  
\sum_{\alpha'''\bk''' \sigma'''}
\sum_{\alpha''\bk'' \sigma''} t_{\alpha'''}^* t_{\alpha''}
\right. \nonumber \\
&&\times
  \leri{
\under{a_{\alpha''' \bk'''  \sigma'''}^{\dagger} c_{\sigma'''}  
\fo{E_{jn}-\sH_0+i\eta} 
a_{\alpha'' \bk''  \sigma''}^{\dagger} c_{\sigma''}}{=\ 0,\  n\text{-2 states}}
+
a_{\alpha''' \bk'''  \sigma'''}^{\dagger} c_{\sigma'''}  
\fo{E_{jn}-\sH_0+i\eta} 
c_{\sigma''}^{\dagger} a_{\alpha'' \bk''  \sigma''} 
\right.  \nonumber \\
&& \left.
+c_{\sigma'''}^{\dagger} a_{\alpha''' \bk'''  \sigma'''} 
\fo{E_{jn}-\sH_0+i\eta} 
a_{\alpha'' \bk''  \sigma''}^{\dagger} c_{\sigma''} 
+\under{c_{\sigma'''}^{\dagger} a_{\alpha''' \bk'''  \sigma'''} 
\fo{E_{jn}-\sH_0+i\eta} 
c_{\sigma''}^{\dagger} a_{\alpha'' \bk''  \sigma''} }{=\ 0,\  n\text{+2 
states}} }
 \nonumber 
 \\
&&\times \left.
 \ket{n}\ket{j}     }^2 \delta(E_{j'n'}-E_{jn}) 
  \nonumber
\end{eqnarray} 
\begin{eqnarray}
\Gamma_{\alpha\alpha'}^{nj;n'j'}
&=& 
\frac{2\pi}{\hbar}\sum_{\bk\bk'\sigma\sigma'}
\ppar{    \bra{j}\bra{n'} a_{\alpha \bk  \sigma}^{\dagger} a_{\alpha' \bk'  
\sigma'}  
\sum_{\alpha''' \bk''' \sigma'''} \sum_{\alpha''\bk'' \sigma''} t_{\alpha'''}^* 
t_{\alpha''} 
\Big\{  
c_{\sigma'''}^{\dagger} a_{\alpha''' \bk'''  \sigma'''} 
\right.   \fo{E_{jn}-\sH_0+i\eta} 
a_{\alpha'' \bk''  \sigma''}^{\dagger} c_{\sigma''}  
 \nonumber \\
&& \left.
+  a_{\alpha''' \bk'''  \sigma'''}^{\dagger} c_{\sigma'''} 
\fo{E_{jn}-\sH_0+i\eta}   
c_{\sigma''}^{\dagger} a_{\alpha'' \bk''  \sigma''} 
\Big\} 
\ket{n}\ket{j}     }^2 \delta(E_{j'n'}-E_{jn}) 
  \nonumber    \\
 &=&\frac{2\pi}{\hbar}\sum_{\bk\bk'\sigma\sigma'}
\ppar{   
\sum_{\alpha''' \bk''' \sigma'''} \sum_{\alpha''\bk'' \sigma''} t_{\alpha'''}^* 
t_{\alpha''} 
\Big\{
\right.  
 \nonumber    \\
 && 
 \bra{j}\bra{n'} a_{\alpha \bk  \sigma}^{\dagger} a_{\alpha' \bk'  \sigma'}  
c_{\sigma'''}^{\dagger} a_{\alpha''' \bk'''  \sigma'''} 
   \fo{E_{jn}-\sH_0+i\eta} 
a_{\alpha'' \bk''  \sigma''}^{\dagger} c_{\sigma''}  \ket{n}\ket{j}   
 \nonumber \\
&& \left.
+ \bra{j}\bra{n'} a_{\alpha \bk  \sigma}^{\dagger} a_{\alpha' \bk'  \sigma'}  
a_{\alpha''' \bk'''  \sigma'''}^{\dagger} c_{\sigma'''} 
\fo{E_{jn}-\sH_0+i\eta}   
c_{\sigma''}^{\dagger} a_{\alpha'' \bk''  \sigma''} 
\ket{n}\ket{j}   
\Big\}    }^2 \delta(E_{j'n'}-E_{jn}) 
  \label{eq:AppGamma}
\end{eqnarray}

Here $n$ and $n'$ are states with the same number of particles. Now we take a 
look at the numerator terms
\begin{eqnarray*}
\bra{j}  a_{\alpha \bk  \sigma}^{\dagger} a_{\alpha' \bk'  \sigma'}  
a_{\alpha''' \bk'''  \sigma'''}a_{\alpha'' \bk''  \sigma''}^{\dagger} \ket{j}  
&=&-
\bra{j}  a_{\alpha \bk  \sigma}^{\dagger}     a_{\alpha''' \bk'''  \sigma'''}
a_{\alpha' \bk'  \sigma'}    a_{\alpha'' \bk''  \sigma''}^{\dagger} \ket{j}
\nonumber \\
&=&-
f\leri{\ve -\mu_{\alpha}} 
\delta_{\alpha \alpha'''}
\delta_{\bk \bk'''}
\delta_{\sigma\sigma'''}
 \nonumber \\
&& \times
\leri{1-f\leri{\ve +\ve_n -\ve_{n'} -\mu_{\alpha'}}} 
\delta_{\alpha'\alpha'' }
\delta_{\bk'  \bk'' }
\delta_{\sigma'\sigma''} 
\end{eqnarray*}

%
%
and

\begin{eqnarray*}
\bra{j} a_{\alpha \bk  \sigma}^{\dagger} a_{\alpha' \bk'  \sigma'}  
   a_{\alpha''' \bk'''  \sigma'''}^{\dagger}  a_{\alpha'' \bk''  \sigma''} 
 \ket{j}&=&
\bra{j} 
a_{\alpha \bk  \sigma}^{\dagger} a_{\alpha' \bk'  \sigma'}  
\leri{\ct{0}{ \delta_{\alpha''' \alpha''}  \delta_{\bk''' \bk''} 
\delta_{\sigma''' \sigma''}} -   a_{\alpha'' \bk''  \sigma''} a_{\alpha''' 
\bk'''  \sigma'''}^{\dagger} }
\ket{j}
 \nonumber \\
&=& 
-\bra{j} 
a_{\alpha \bk  \sigma}^{\dagger} a_{\alpha' \bk'  \sigma'}  
a_{\alpha'' \bk''  \sigma''} a_{\alpha''' \bk'''  \sigma'''}^{\dagger} 
\ket{j}
 \nonumber \\
&=& 
\bra{j} 
a_{\alpha \bk  \sigma}^{\dagger}  a_{\alpha'' \bk''  \sigma''}
\ket{j}\bra{j}
a_{\alpha' \bk'  \sigma'}  a_{\alpha''' \bk'''  \sigma'''}^{\dagger} 
\ket{j}
 \nonumber \\
&=& 
f\leri{\ve -\mu_{\alpha}} 
\delta_{\alpha \alpha''}
\delta_{\bk \bk''}
\delta_{\sigma\sigma''}
 \nonumber \\
&& 
\leri{1-f\leri{\ve +\ve_n -\ve_{n'} -\mu_{\alpha'}}} 
\delta_{\alpha'\alpha''' }
\delta_{\bk'  \bk''' }
\delta_{\sigma'\sigma'''}  
\end{eqnarray*}

Here we have used a Taylor series expansion on the operator 
$1/(E_{jn}-H_0)=(1/E_{jn})\sum_{l=0}^{\infty}(H_0/E_{jn})^l$.

Taking into account last delta rules, we have
\begin{eqnarray*}
\bra{n'} c^{\dagger}_{\sigma'''}c_{\sigma''} \ket{n}&=&
\sum_{n''}\bra{n'} c^{\dagger}_{\sigma} \ket{n''}  \bra{n''}c_{\sigma'} \ket{n}
=\sum_{n''} (\bra{n''} c_{\sigma} \ket{n'})^{\dagger}  \bra{n''}c_{\sigma'} 
\ket{n}
=\sum_{n''} A_{n''n'}^{\sigma *} 
           A_{n''n}^{\sigma'}
\end{eqnarray*}
and 
\begin{eqnarray*}
\bra{n'} c_{\sigma'}c_{\sigma}^{\dagger} \ket{n}&=&
\sum_{n''}\bra{n'} c_{\sigma'} \ket{n''}  \bra{n''}c_{\sigma}^{\dagger} \ket{n}
=\sum_{n''}\bra{n'} c_{\sigma'} \ket{n''}  
\leri{\bra{n}c_{\sigma}\ket{n''}}^{\dagger}
=\sum_{n''} A_{n'n''}^{\sigma'} 
           A_{nn''}^{\sigma *}
\end{eqnarray*}
where $A_{n'n''}^{\sigma'} =\bra{n'} c_{\sigma'} \ket{n''} $ and 
$A_{nn''}^{\sigma *}=\bra{n''}c_{\sigma}^{\dagger} \ket{n}$. Here $n''$ 
represents a intermediate state.

Thus Eq. (\ref{eq:AppGamma}) becomes

\begin{eqnarray}
\Gamma_{\alpha\alpha'}^{nj;n'j'}
 &=&\frac{2\pi}{\hbar}\sum_{\bk\bk'\sigma\sigma'}
\ppar{   
\sum_{\alpha''' \bk''' \sigma'''} \sum_{\alpha''\bk'' \sigma''} t_{\alpha'''}^* 
t_{\alpha''} 
\Big\{
\right.  
 \nonumber    \\
 && 
 -\bra{j}\bra{n'} a_{\alpha \bk  \sigma}^{\dagger} a_{\alpha' \bk'  \sigma'}   
a_{\alpha''' \bk'''  \sigma'''} 
   c_{\sigma'''}^{\dagger} \fo{\ve_{n'}-\ve_{n''}-\ve+i\eta} 
   c_{\sigma''}  \ket{n} a_{\alpha'' \bk''  \sigma''}^{\dagger} \ket{j}   
 \nonumber \\
&& \left.
+ \bra{j}\bra{n'} a_{\alpha \bk  \sigma}^{\dagger} a_{\alpha' \bk'  \sigma'}  
a_{\alpha''' \bk'''  \sigma'''}^{\dagger} c_{\sigma'''} 
\fo{\ve_{n}-\ve_{n''}+\ve+i\eta}   
c_{\sigma''}^{\dagger} \ket{n} a_{\alpha'' \bk''  \sigma''} 
\ket{j}   
\Big\}    }^2 \delta(E_{j'n'}-E_{jn}) 
  \nonumber 
  \\
\Gamma_{\alpha\alpha'}^{n;n'} &=&2
\ppar{t_{\alpha}}^2 \ppar{t_{\alpha'}}^2 \sum_{\sigma\sigma'}  
\nu_{\alpha}(\sigma) \nu_{\alpha'}(\sigma')
\int \de\ve f\leri{\ve -\mu_{\alpha}} 
\leri{1-f\leri{\ve +\ve_n -\ve_{n'} -\mu_{\alpha'}}}
  \nonumber 
  \\
 &&\ \ \ \ \ \ \ \ \ \ \ \ \ \ \ \ \ \  \ \ \ \ \ \ \ \ \ \  \times
\ppar{
\sum_{n''} 
\pbk{
\frac{ A_{n''n'}^{\sigma *} A_{n''n}^{\sigma'}}
{\ve-\ve_{n'}+\ve_{n''}+i\eta}  +
\frac{ A_{n'n''}^{\sigma'}A_{nn''}^{\sigma *}}
{\ve+\ve_{n}-\ve_{n''}+i\eta}} }^2
  \nonumber \\
 &=&
\sum_{\sigma\sigma'} \gamma_{\alpha}^{\sigma}\gamma_{\alpha'}^{\sigma'}
\int \de\ve 
 f\leri{\ve -\mu_{\alpha}} 
\leri{1-f\leri{\ve +\ve_n -\ve_{n'} -\mu_{\alpha'}}}
   \nonumber 
  \\
 &&\ \ \ \ \ \ \ \ \ \ \ \ \ \ \ \ \ \  \ \ \ \ \ \ \ \ \ \ \times
\under{
\ppar{
\sum_{n''} 
\pbk{
\frac{ A_{n''n'}^{\sigma *} A_{n''n}^{\sigma'}}
{\ve-\ve_{n'}+\ve_{n''}+i\eta}  +
\frac{ A_{n'n''}^{\sigma'}A_{nn''}^{\sigma *}}
{\ve+\ve_{n}-\ve_{n''}+i\eta}} }^2 
}{{Q}}
 \label{eq:Appgammasum}
\end{eqnarray}

%
%
%
%
%
%
%
%

\section{\label{sec:rightrates} Explicit derivation of 
Eq.~(\ref{eq:cotunnelingrates})}

The absolute value in Eq. (\ref{eq:Appgammasum}) can be written as
\begin{eqnarray*}
{Q} &=&\ppar{
\sum_{n''} 
\pbk{
\frac{ A_{n''n'}^{\sigma *} A_{n''n}^{\sigma'}}
{\ve-\ve_{n'}+\ve_{n''}+i\eta}  +
\frac{ A_{n'n''}^{\sigma'}A_{nn''}^{\sigma *}}
{\ve+\ve_{n}-\ve_{n''}+i\eta} }}^2 \nonumber 
\\
&=&\leri{
\frac{  A_{1n}^{\sigma'*}A_{1n'}^{\sigma }}
{\ve-\ve_{n'}+\ve_{1}-i\eta}  +
\frac{ A_{n1}^{\sigma }A_{n'1}^{\sigma'*}}
{\ve+\ve_{n}-\ve_{1}-i\eta}
+
\frac{  A_{2n}^{\sigma'*}A_{2n'}^{\sigma}}
{\ve-\ve_{n'}+\ve_{2}-i\eta}  +
\frac{ A_{n2}^{\sigma }A_{n'2}^{\sigma'*}}
{\ve+\ve_{n}-\ve_{2}-i\eta} 
\right.
\nonumber 
\\
&&+\left.
\frac{ A_{3n}^{\sigma'*}A_{3n'}^{\sigma }}
{\ve-\ve_{n'}+\ve_{n''}-i\eta}  +
\frac{ A_{n3}^{\sigma }A_{n'3}^{\sigma'*}}
{\ve+\ve_{n}-\ve_{3}-i\eta}}
\nonumber 
\\
&&\times\leri{
\frac{ A_{1n'}^{\sigma *} A_{1n}^{\sigma'}}
{\ve-\ve_{n'}+\ve_{1}+i\eta}  +
\frac{ A_{n'1}^{\sigma'}A_{n1}^{\sigma *}}
{\ve+\ve_{n}-\ve_{1}+i\eta}
+
\frac{ A_{2n'}^{\sigma *} A_{2n}^{\sigma'}}
{\ve-\ve_{n'}+\ve_{2}+i\eta}  +
\frac{ A_{n'2}^{\sigma'}A_{n2}^{\sigma *}}
{\ve+\ve_{n}-\ve_{2}+i\eta} 
\right.
\nonumber 
\\
&&+\left.
\frac{ A_{3n'}^{\sigma *} A_{3n}^{\sigma'}}
{\ve-\ve_{n'}+\ve_{n''}+i\eta}  +
\frac{ A_{n'3}^{\sigma'}A_{n3}^{\sigma *}}
{\ve+\ve_{n}-\ve_{3}+i\eta}}\nonumber 
\end{eqnarray*}

\begin{eqnarray*}
{Q}
&=&
\sum_k\leri{
\frac{ (A_{kn'}^{\sigma *} A_{kn}^{\sigma'})^2}
{(\ve-\ve_{n'}+\ve_{k})^2+\eta^2}  +
\frac{ (A_{n'k}^{\sigma'}A_{nk}^{\sigma *})^2}
{(\ve+\ve_{n}-\ve_{k})^2+\eta^2}}  \nonumber \\
&&+2\Real
\sum_q\sum_{k<q}\leri{ 
\frac{ A_{qn'}^{\sigma *} A_{qn}^{\sigma'}}
{\ve-\ve_{n'}+\ve_{q}+i\eta}  
\frac{ A_{kn'}^{\sigma *} A_{kn}^{\sigma'}}
{\ve-\ve_{n'}+\ve_{k}-i\eta} 
+
\frac{ A_{n'q}^{\sigma'}A_{nq}^{\sigma *}}
{\ve+\ve_{n}-\ve_{q}+i\eta}  
\frac{ A_{n'k}^{\sigma'}A_{nk}^{\sigma *}}
{\ve+\ve_{n}-\ve_{k}-i\eta} 
}
\nonumber 
\\
&&+2\Real
\sum_q\sum_{k}\leri{ 
\frac{ A_{kn'}^{\sigma *} A_{kn}^{\sigma'}}
{\ve-\ve_{n'}+\ve_{q}-i\eta} 
\frac{ A_{n'k}^{\sigma'}A_{nk}^{\sigma *}}
{\ve+\ve_{n}-\ve_{k}-i\eta} 
}
\end{eqnarray*}
Thus Eq.~(\ref{eq:gammasum}) becomes 
\begin{eqnarray*}
\Gamma_{\alpha\alpha'}^{n;n'}
&=&
\sum_{\sigma\sigma'} \gamma_{\alpha}^{\sigma}\gamma_{\alpha'}^{\sigma'}
\int \de\ve 
 f\leri{\ve -\mu_{\alpha}} 
\leri{1-f\leri{\ve +\ve_n -\ve_{n'} -\mu_{\alpha'}}}
   \nonumber 
  \\
 &&\times
\ppar{
\sum_{n''} 
\pbk{
\frac{ A_{n''n'}^{\sigma *} A_{n''n}^{\sigma'}}
{\ve-\ve_{n'}+\ve_{n''}+i\eta}  +
\frac{ A_{n'n''}^{\sigma'}A_{nn''}^{\sigma *}}
{\ve+\ve_{n}-\ve_{n''}+i\eta}} }^2
  \nonumber 
  \\
 &=& 
 \sum_{\sigma\sigma'} \gamma_{\alpha}^{\sigma}\gamma_{\alpha'}^{\sigma'}
\int \de\ve 
 f\leri{\ve -\mu_{\alpha}} 
\leri{1-f\leri{\ve +\ve_n -\ve_{n'} -\mu_{\alpha'}}}
   \nonumber 
  \\
 && \times  \psqrt{
\sum_k\leri{
\frac{ (A_{kn'}^{\sigma *} A_{kn}^{\sigma'})^2}
{(\ve-\ve_{n'}+\ve_{k})^2+\eta^2}  +
\frac{ (A_{n'k}^{\sigma'}A_{nk}^{\sigma *})^2}
{(\ve+\ve_{n}-\ve_{k})^2+\eta^2}}  \right. \nonumber \\
&&+2\Real
\sum_q\sum_{k<q}\leri{ 
\frac{ A_{qn'}^{\sigma *} A_{qn}^{\sigma'}}
{\ve-\ve_{n'}+\ve_{q}+i\eta}  
\frac{ A_{kn'}^{\sigma *} A_{kn}^{\sigma'}}
{\ve-\ve_{n'}+\ve_{k}-i\eta} 
+
\frac{ A_{n'q}^{\sigma'}A_{nq}^{\sigma *}}
{\ve+\ve_{n}-\ve_{q}+i\eta}  
\frac{ A_{n'k}^{\sigma'}A_{nk}^{\sigma *}}
{\ve+\ve_{n}-\ve_{k}-i\eta} 
}
\nonumber 
\\
&&+\left.
2\Real
\sum_q\sum_{k}\leri{ 
\frac{ A_{kn'}^{\sigma *} A_{kn}^{\sigma'}}
{\ve-\ve_{n'}+\ve_{q}-i\eta} 
\frac{ A_{n'k}^{\sigma'}A_{nk}^{\sigma *}}
{\ve+\ve_{n}-\ve_{k}-i\eta} 
} }
  \nonumber 
 \end{eqnarray*}

\begin{eqnarray*}
\Gamma_{\alpha\alpha'}^{n;n'}
&=&
\sum_{\sigma\sigma'} \gamma_{\alpha}^{\sigma}\gamma_{\alpha'}^{\sigma'}
\int \de\ve 
 f\leri{\ve -E_1} 
\leri{1-f\leri{\ve -E_2}}
   \nonumber 
  \\
 && \times  \psqrt{
\sum_k
\frac{ A^2}
{(\ve-\ve_{ak})^2+\eta^2}  \ \ \ \ \ \ \ \ \ \ \  {\text{(Integral type 
J)  }}
 \right. \nonumber \\
&&
+\sum_k
\frac{ B^2}
{(\ve-\ve_{bk})^2+\eta^2}  \ \ \ \ \ \ \ \ \ \ \  {\text{(Integral type 
J)  }}  \nonumber \\
&&+2\Real
\sum_q\sum_{k<q}
\frac{ A_k}
{\ve-\ve_{ak}+i\eta}  
\frac{ A_{q}}
{\ve-\ve_{aq}-i\eta}    \ \ \ \ \ \ \ \ \ \ \  {\text{(Integral type I)}  
}
  \nonumber \\
&&+2\Real
\sum_q\sum_{k<q}
\frac{ B_{k}}
{\ve-\ve_{bk}+i\eta}  
\frac{ B_{q}}
{\ve-\ve_{bq}-i\eta}  \ \ \ \ \ \ \ \ \ \ \  {\text{(Integral type I)  }}
\nonumber 
\\
&&+\left.
2\Real
\sum_q\sum_{k}
\frac{ A_k}
{\ve-\ve_{ak}+i\eta} 
\frac{ B_{q}}
{\ve-\ve_{bq}-i\eta} 
 }    \ \ \ \ \ \ \ \ \ \ \  {\text{(Integral type I)  }}
\label{eq:integrals}
\end{eqnarray*}
where $A_k=A_{kn'}^{\sigma *} A_{kn}^{\sigma'}$, 
$B_k=A_{n'k}^{\sigma'}A_{nk}^{\sigma *}$, $\ve_{ak}=\ve_{n'}-\ve_{k}$,  
$\ve_{bk}=\ve_{k}-\ve_{n}$, $E_1=\mu_{\alpha}$ and  
$E_2=\mu_{\alpha'}+\ve_{n'}-\ve_{n}$.
\vspace{1cm}

{Integral type I}
\begin{eqnarray}
I(E_1,E_2,\ve_1,\ve_2)&=& \mathfrak{Re} \int \de \ve f(\ve 
-E_1)\psqrt{1-f(\ve-E_2)}
\fo{\ve-\ve_1-i\gamma}\fo{\ve-\ve_2+i\gamma}  
\nonumber \\
&=&\f{n_{B}(E_2 -E_1)}{\ve_1 -\ve_2 } \Real \pbk{   
 \psi\leri{\f{1}{2} + \f{i\beta}{2\pi }\psqrt{E_2-\ve_1}}
-\psi\leri{\f{1}{2} - \f{i\beta}{2\pi }\psqrt{E_2-\ve_2}} \right.
\nonumber \\
&&\left.
\ \ \ \ \ \ \ \ \ \ \ \ \ -\psi\leri{\f{1}{2} + \f{i\beta}{2\pi 
}\psqrt{E_1-\ve_1}}  
+\psi\leri{\f{1}{2} - \f{i\beta}{2\pi }\psqrt{E_1-\ve_2}} }
\label{eq:I}
\end{eqnarray}
Here $\psi$ is the digamma function, $n_B$ is the Bose function and $\beta = 
1/k_BT$.

\vspace{1cm}

{Integral type J}
\begin{eqnarray}
J(E_1,E_2,\ve_1)&=&\int \de \ve f(\ve-E_1)\left[1-f(\ve-E_2)\right]\f{1}{(\ve - 
\ve_1)^2 + \eta^2} \nonumber \\
&=&\f{\beta }{2\pi }n_{B}(E_2 -E_1) \Ima \pbk{   
 \psi'\leri{\f{1}{2} + \f{i\beta}{2\pi }\psqrt{E_2-\ve_1}}
-\psi'\leri{\f{1}{2} + \f{i\beta}{2\pi }\psqrt{E_1-\ve_1}} }
\label{eq:J}
\end{eqnarray}

Thus Eq. (\ref{eq:integrals}) becomes

\begin{eqnarray}
\Gamma_{\alpha\alpha'}^{n;n'}
&=& 
\sum_{\sigma\sigma'} \gamma_{\alpha}^{\sigma}\gamma_{\alpha'}^{\sigma'}
 \psqrt{
\sum_k  \leri{
 A^2 J(E_1,E_2,\ve_{ak})
+B^2 J(E_1,E_2,\ve_{bk} )  }
\right. \nonumber \\
&&+2
\sum_q\sum_{k\neq q}   \leri{
   A_k A_{q} I(E_1,E_2,\ve_{ak},\ve_{aq}) 
+ B_{k}B_{q} I(E_1,E_2, \ve_{bk},\ve_{bq})}
\nonumber \\
&&+\left.	
2\sum_q\sum_{k}
A_k B_{q} I(E_1,E_2,\ve_{ak},\ve_{bq}) }
\label{eq:appcotunnelingrates}  
\end{eqnarray}

\twocolumngrid

\bibliography{SEtransport}

\providecommand{\noopsort}[1]{}\providecommand{\singleletter}[1]{#1}%
\begin{thebibliography}{64}%
\makeatletter
\providecommand \@ifxundefined [1]{%
 \@ifx{#1\undefined}
}%
\providecommand \@ifnum [1]{%
 \ifnum #1\expandafter \@firstoftwo
 \else \expandafter \@secondoftwo
 \fi
}%
\providecommand \@ifx [1]{%
 \ifx #1\expandafter \@firstoftwo
 \else \expandafter \@secondoftwo
 \fi
}%
\providecommand \natexlab [1]{#1}%
\providecommand \enquote  [1]{``#1''}%
\providecommand \bibnamefont  [1]{#1}%
\providecommand \bibfnamefont [1]{#1}%
\providecommand \citenamefont [1]{#1}%
\providecommand \href@noop [0]{\@secondoftwo}%
\providecommand \href [0]{\begingroup \@sanitize@url \@href}%
\providecommand \@href[1]{\@@startlink{#1}\@@href}%
\providecommand \@@href[1]{\endgroup#1\@@endlink}%
\providecommand \@sanitize@url [0]{\catcode `\\12\catcode `\$12\catcode
  `\&12\catcode `\#12\catcode `\^12\catcode `\_12\catcode `\%12\relax}%
\providecommand \@@startlink[1]{}%
\providecommand \@@endlink[0]{}%
\providecommand \url  [0]{\begingroup\@sanitize@url \@url }%
\providecommand \@url [1]{\endgroup\@href {#1}{\urlprefix }}%
\providecommand \urlprefix  [0]{URL }%
\providecommand \Eprint [0]{\href }%
\providecommand \doibase [0]{http://dx.doi.org/}%
\providecommand \selectlanguage [0]{\@gobble}%
\providecommand \bibinfo  [0]{\@secondoftwo}%
\providecommand \bibfield  [0]{\@secondoftwo}%
\providecommand \translation [1]{[#1]}%
\providecommand \BibitemOpen [0]{}%
\providecommand \bibitemStop [0]{}%
\providecommand \bibitemNoStop [0]{.\EOS\space}%
\providecommand \EOS [0]{\spacefactor3000\relax}%
\providecommand \BibitemShut  [1]{\csname bibitem#1\endcsname}%
\let\auto@bib@innerbib\@empty
\bibitem [{\citenamefont {Gatteschi}\ \emph {et~al.}(2006)\citenamefont
  {Gatteschi}, \citenamefont {Sessoli},\ and\ \citenamefont
  {Villain}}]{Gatteschi2006}%
  \BibitemOpen
  \bibfield  {author} {\bibinfo {author} {\bibfnamefont {D.}~\bibnamefont
  {Gatteschi}}, \bibinfo {author} {\bibfnamefont {R.}~\bibnamefont {Sessoli}},
  \ and\ \bibinfo {author} {\bibfnamefont {J.}~\bibnamefont {Villain}},\
  }\href@noop {} {\emph {\bibinfo {title} {Molecular Nanomagnets}}}\ (\bibinfo
  {publisher} {Oxford University Press},\ \bibinfo {address} {Oxford},\
  \bibinfo {year} {2006})\BibitemShut {NoStop}%
\bibitem [{\citenamefont {Wedge}\ \emph {et~al.}(2012)\citenamefont {Wedge},
  \citenamefont {Timco}, \citenamefont {Spielberg}, \citenamefont {George},
  \citenamefont {Tuna}, \citenamefont {Rigby}, \citenamefont {McInnes},
  \citenamefont {Winpenny}, \citenamefont {Blundell},\ and\ \citenamefont
  {Ardavan}}]{Wedge2012}%
  \BibitemOpen
  \bibfield  {author} {\bibinfo {author} {\bibfnamefont {C.~J.}\ \bibnamefont
  {Wedge}}, \bibinfo {author} {\bibfnamefont {G.~A.}\ \bibnamefont {Timco}},
  \bibinfo {author} {\bibfnamefont {E.~T.}\ \bibnamefont {Spielberg}}, \bibinfo
  {author} {\bibfnamefont {R.~E.}\ \bibnamefont {George}}, \bibinfo {author}
  {\bibfnamefont {F.}~\bibnamefont {Tuna}}, \bibinfo {author} {\bibfnamefont
  {S.}~\bibnamefont {Rigby}}, \bibinfo {author} {\bibfnamefont {E.~J.~L.}\
  \bibnamefont {McInnes}}, \bibinfo {author} {\bibfnamefont {R.~E.~P.}\
  \bibnamefont {Winpenny}}, \bibinfo {author} {\bibfnamefont {S.~J.}\
  \bibnamefont {Blundell}}, \ and\ \bibinfo {author} {\bibfnamefont
  {A.}~\bibnamefont {Ardavan}},\ }\href
  {http://link.aps.org/doi/10.1103/PhysRevLett.108.107204} {\bibfield
  {journal} {\bibinfo  {journal} {Phys. Rev. Lett.}\ }\textbf {\bibinfo
  {volume} {108}},\ \bibinfo {pages} {107204} (\bibinfo {year}
  {2012})}\BibitemShut {NoStop}%
\bibitem [{\citenamefont {Bogani}\ and\ \citenamefont
  {Wernsdorfer}(2008)}]{Bogani2008}%
  \BibitemOpen
  \bibfield  {author} {\bibinfo {author} {\bibfnamefont {L.}~\bibnamefont
  {Bogani}}\ and\ \bibinfo {author} {\bibfnamefont {W.}~\bibnamefont
  {Wernsdorfer}},\ }\href@noop {} {\bibfield  {journal} {\bibinfo  {journal}
  {Nat Mater}\ }\textbf {\bibinfo {volume} {7}},\ \bibinfo {pages} {179}
  (\bibinfo {year} {2008})}\BibitemShut {NoStop}%
\bibitem [{\citenamefont {Sanvito}(2011)}]{Sanvito2011}%
  \BibitemOpen
  \bibfield  {author} {\bibinfo {author} {\bibfnamefont {S.}~\bibnamefont
  {Sanvito}},\ }\href@noop {} {\bibfield  {journal} {\bibinfo  {journal} {Chem.
  Soc. Rev.}\ }\textbf {\bibinfo {volume} {40}},\ \bibinfo {pages} {3336}
  (\bibinfo {year} {2011})}\BibitemShut {NoStop}%
\bibitem [{\citenamefont {Affronte}(2009)}]{Affronte2009}%
  \BibitemOpen
  \bibfield  {author} {\bibinfo {author} {\bibfnamefont {M.}~\bibnamefont
  {Affronte}},\ }\href@noop {} {\bibfield  {journal} {\bibinfo  {journal} {J.
  Mater. Chem.}\ }\textbf {\bibinfo {volume} {19}},\ \bibinfo {pages} {1731}
  (\bibinfo {year} {2009})}\BibitemShut {NoStop}%
\bibitem [{\citenamefont {Leuenberger}\ and\ \citenamefont
  {Loss}(2001)}]{Leuenberger2001}%
  \BibitemOpen
  \bibfield  {author} {\bibinfo {author} {\bibfnamefont {M.~N.}\ \bibnamefont
  {Leuenberger}}\ and\ \bibinfo {author} {\bibfnamefont {D.}~\bibnamefont
  {Loss}},\ }\href@noop {} {\bibfield  {journal} {\bibinfo  {journal} {Nature}\
  }\textbf {\bibinfo {volume} {410}},\ \bibinfo {pages} {789} (\bibinfo {year}
  {2001})}\BibitemShut {NoStop}%
\bibitem [{\citenamefont {Lehmann}\ \emph {et~al.}(2007)\citenamefont
  {Lehmann}, \citenamefont {Gaita-Arino}, \citenamefont {Coronado},\ and\
  \citenamefont {Loss}}]{Lehmann2007}%
  \BibitemOpen
  \bibfield  {author} {\bibinfo {author} {\bibfnamefont {J.}~\bibnamefont
  {Lehmann}}, \bibinfo {author} {\bibfnamefont {A.}~\bibnamefont
  {Gaita-Arino}}, \bibinfo {author} {\bibfnamefont {E.}~\bibnamefont
  {Coronado}}, \ and\ \bibinfo {author} {\bibfnamefont {D.}~\bibnamefont
  {Loss}},\ }\href@noop {} {\bibfield  {journal} {\bibinfo  {journal} {Nat
  Nano}\ }\textbf {\bibinfo {volume} {2}},\ \bibinfo {pages} {312} (\bibinfo
  {year} {2007})}\BibitemShut {NoStop}%
\bibitem [{\citenamefont {Ardavan}\ \emph {et~al.}(2007)\citenamefont
  {Ardavan}, \citenamefont {Rival}, \citenamefont {Morton}, \citenamefont
  {Blundell}, \citenamefont {Tyryshkin}, \citenamefont {Timco},\ and\
  \citenamefont {Winpenny}}]{Ardavan2007}%
  \BibitemOpen
  \bibfield  {author} {\bibinfo {author} {\bibfnamefont {A.}~\bibnamefont
  {Ardavan}}, \bibinfo {author} {\bibfnamefont {O.}~\bibnamefont {Rival}},
  \bibinfo {author} {\bibfnamefont {J.~J.~L.}\ \bibnamefont {Morton}}, \bibinfo
  {author} {\bibfnamefont {S.~J.}\ \bibnamefont {Blundell}}, \bibinfo {author}
  {\bibfnamefont {A.~M.}\ \bibnamefont {Tyryshkin}}, \bibinfo {author}
  {\bibfnamefont {G.~A.}\ \bibnamefont {Timco}}, \ and\ \bibinfo {author}
  {\bibfnamefont {R.~E.~P.}\ \bibnamefont {Winpenny}},\ }\href@noop {}
  {\bibfield  {journal} {\bibinfo  {journal} {Phys. Rev. Lett.}\ }\textbf
  {\bibinfo {volume} {98}},\ \bibinfo {pages} {057201} (\bibinfo {year}
  {2007})}\BibitemShut {NoStop}%
\bibitem [{\citenamefont {Andre}\ \emph {et~al.}(2006)\citenamefont {Andre},
  \citenamefont {DeMille}, \citenamefont {Doyle}, \citenamefont {Lukin},
  \citenamefont {Maxwell}, \citenamefont {Rabl}, \citenamefont {Schoelkopf},\
  and\ \citenamefont {Zoller}}]{Andre2006}%
  \BibitemOpen
  \bibfield  {author} {\bibinfo {author} {\bibfnamefont {A.}~\bibnamefont
  {Andre}}, \bibinfo {author} {\bibfnamefont {D.}~\bibnamefont {DeMille}},
  \bibinfo {author} {\bibfnamefont {J.~M.}\ \bibnamefont {Doyle}}, \bibinfo
  {author} {\bibfnamefont {M.~D.}\ \bibnamefont {Lukin}}, \bibinfo {author}
  {\bibfnamefont {S.~E.}\ \bibnamefont {Maxwell}}, \bibinfo {author}
  {\bibfnamefont {P.}~\bibnamefont {Rabl}}, \bibinfo {author} {\bibfnamefont
  {R.~J.}\ \bibnamefont {Schoelkopf}}, \ and\ \bibinfo {author} {\bibfnamefont
  {P.}~\bibnamefont {Zoller}},\ }\href {http://dx.doi.org/10.1038/nphys386}
  {\bibfield  {journal} {\bibinfo  {journal} {Nat Phys}\ }\textbf {\bibinfo
  {volume} {2}},\ \bibinfo {pages} {636} (\bibinfo {year} {2006})}\BibitemShut
  {NoStop}%
\bibitem [{\citenamefont {Hirjibehedin}\ \emph {et~al.}(2006)\citenamefont
  {Hirjibehedin}, \citenamefont {Lutz},\ and\ \citenamefont
  {Heinrich}}]{Hirjibehedin2006}%
  \BibitemOpen
  \bibfield  {author} {\bibinfo {author} {\bibfnamefont {C.~F.}\ \bibnamefont
  {Hirjibehedin}}, \bibinfo {author} {\bibfnamefont {C.~P.}\ \bibnamefont
  {Lutz}}, \ and\ \bibinfo {author} {\bibfnamefont {A.~J.}\ \bibnamefont
  {Heinrich}},\ }\href@noop {} {\bibfield  {journal} {\bibinfo  {journal}
  {Science}\ }\textbf {\bibinfo {volume} {312}},\ \bibinfo {pages} {1021}
  (\bibinfo {year} {2006})}\BibitemShut {NoStop}%
\bibitem [{\citenamefont {Bleszynski-Jayich}\ \emph {et~al.}(2008)\citenamefont
  {Bleszynski-Jayich}, \citenamefont {Fr�berg}, \citenamefont {Bj�rk},
  \citenamefont {Trodahl}, \citenamefont {Samuelson},\ and\ \citenamefont
  {Westervelt}}]{Bleszynski-Jayich2008}%
  \BibitemOpen
  \bibfield  {author} {\bibinfo {author} {\bibfnamefont {A.~C.}\ \bibnamefont
  {Bleszynski-Jayich}}, \bibinfo {author} {\bibfnamefont {L.~E.}\ \bibnamefont
  {Fr�berg}}, \bibinfo {author} {\bibfnamefont {M.~T.}\ \bibnamefont
  {Bj�rk}}, \bibinfo {author} {\bibfnamefont {H.~J.}\ \bibnamefont
  {Trodahl}}, \bibinfo {author} {\bibfnamefont {L.}~\bibnamefont {Samuelson}},
  \ and\ \bibinfo {author} {\bibfnamefont {R.~M.}\ \bibnamefont {Westervelt}},\
  }\href@noop {} {\bibfield  {journal} {\bibinfo  {journal} {Phys. Rev. B}\
  }\textbf {\bibinfo {volume} {77}},\ \bibinfo {pages} {245327} (\bibinfo
  {year} {2008})}\BibitemShut {NoStop}%
\bibitem [{\citenamefont {Nowack}\ \emph {et~al.}(2007)\citenamefont {Nowack},
  \citenamefont {Koppens}, \citenamefont {Nazarov},\ and\ \citenamefont
  {Vandersypen}}]{kowack2007}%
  \BibitemOpen
  \bibfield  {author} {\bibinfo {author} {\bibfnamefont {K.~C.}\ \bibnamefont
  {Nowack}}, \bibinfo {author} {\bibfnamefont {F.~H.~L.}\ \bibnamefont
  {Koppens}}, \bibinfo {author} {\bibfnamefont {Y.~V.}\ \bibnamefont
  {Nazarov}}, \ and\ \bibinfo {author} {\bibfnamefont {L.~M.~K.}\ \bibnamefont
  {Vandersypen}},\ }\href@noop {} {\bibfield  {journal} {\bibinfo  {journal}
  {Science}\ }\textbf {\bibinfo {volume} {318}},\ \bibinfo {pages} {1430}
  (\bibinfo {year} {2007})}\BibitemShut {NoStop}%
\bibitem [{\citenamefont {Baadji}\ \emph {et~al.}(2009)\citenamefont {Baadji},
  \citenamefont {Piacenza}, \citenamefont {Tugsuz}, \citenamefont {Sala},
  \citenamefont {Maruccio},\ and\ \citenamefont
  {Sanvito}}]{sanvito_nat_mat_2009}%
  \BibitemOpen
  \bibfield  {author} {\bibinfo {author} {\bibfnamefont {N.}~\bibnamefont
  {Baadji}}, \bibinfo {author} {\bibfnamefont {M.}~\bibnamefont {Piacenza}},
  \bibinfo {author} {\bibfnamefont {T.}~\bibnamefont {Tugsuz}}, \bibinfo
  {author} {\bibfnamefont {F.~D.}\ \bibnamefont {Sala}}, \bibinfo {author}
  {\bibfnamefont {G.}~\bibnamefont {Maruccio}}, \ and\ \bibinfo {author}
  {\bibfnamefont {S.}~\bibnamefont {Sanvito}},\ }\href@noop {} {\bibfield
  {journal} {\bibinfo  {journal} {Nat. Mat.}\ }\textbf {\bibinfo {volume}
  {8}},\ \bibinfo {pages} {813} (\bibinfo {year} {2009})}\BibitemShut {NoStop}%
\bibitem [{\citenamefont {Osorio}\ \emph {et~al.}(2010)\citenamefont {Osorio},
  \citenamefont {Moth-Poulsen}, \citenamefont {van~der Zant}, \citenamefont
  {Paaske}, \citenamefont {Hedeg{\aa}rd}, \citenamefont {Flensberg},
  \citenamefont {Bendix},\ and\ \citenamefont
  {Bj{\o}rnholm}}]{van_der_zant_2010}%
  \BibitemOpen
  \bibfield  {author} {\bibinfo {author} {\bibfnamefont {E.~A.}\ \bibnamefont
  {Osorio}}, \bibinfo {author} {\bibfnamefont {K.}~\bibnamefont
  {Moth-Poulsen}}, \bibinfo {author} {\bibfnamefont {H.~S.~J.}\ \bibnamefont
  {van~der Zant}}, \bibinfo {author} {\bibfnamefont {J.}~\bibnamefont
  {Paaske}}, \bibinfo {author} {\bibfnamefont {P.}~\bibnamefont
  {Hedeg{\aa}rd}}, \bibinfo {author} {\bibfnamefont {K.}~\bibnamefont
  {Flensberg}}, \bibinfo {author} {\bibfnamefont {J.}~\bibnamefont {Bendix}}, \
  and\ \bibinfo {author} {\bibfnamefont {T.}~\bibnamefont {Bj{\o}rnholm}},\
  }\href@noop {} {\bibfield  {journal} {\bibinfo  {journal} {Nanolett.}\
  }\textbf {\bibinfo {volume} {10}},\ \bibinfo {pages} {105} (\bibinfo {year}
  {2010})}\BibitemShut {NoStop}%
\bibitem [{\citenamefont {Trif}\ \emph {et~al.}(2008)\citenamefont {Trif},
  \citenamefont {Troiani}, \citenamefont {Stepanenko},\ and\ \citenamefont
  {Loss}}]{Trif2008}%
  \BibitemOpen
  \bibfield  {author} {\bibinfo {author} {\bibfnamefont {M.}~\bibnamefont
  {Trif}}, \bibinfo {author} {\bibfnamefont {F.}~\bibnamefont {Troiani}},
  \bibinfo {author} {\bibfnamefont {D.}~\bibnamefont {Stepanenko}}, \ and\
  \bibinfo {author} {\bibfnamefont {D.}~\bibnamefont {Loss}},\ }\href@noop {}
  {\bibfield  {journal} {\bibinfo  {journal} {Phys. Rev. Lett.}\ }\textbf
  {\bibinfo {volume} {101}},\ \bibinfo {pages} {217201} (\bibinfo {year}
  {2008})}\BibitemShut {NoStop}%
\bibitem [{\citenamefont {Choi}\ \emph {et~al.}(2006)\citenamefont {Choi},
  \citenamefont {Matsuda}, \citenamefont {Nojiri}, \citenamefont {Kortz},
  \citenamefont {Hussain}, \citenamefont {Stowe}, \citenamefont {Ramsey},\ and\
  \citenamefont {Dalal}}]{Choi2006}%
  \BibitemOpen
  \bibfield  {author} {\bibinfo {author} {\bibfnamefont {K.-Y.}\ \bibnamefont
  {Choi}}, \bibinfo {author} {\bibfnamefont {Y.~H.}\ \bibnamefont {Matsuda}},
  \bibinfo {author} {\bibfnamefont {H.}~\bibnamefont {Nojiri}}, \bibinfo
  {author} {\bibfnamefont {U.}~\bibnamefont {Kortz}}, \bibinfo {author}
  {\bibfnamefont {F.}~\bibnamefont {Hussain}}, \bibinfo {author} {\bibfnamefont
  {A.~C.}\ \bibnamefont {Stowe}}, \bibinfo {author} {\bibfnamefont
  {C.}~\bibnamefont {Ramsey}}, \ and\ \bibinfo {author} {\bibfnamefont {N.~S.}\
  \bibnamefont {Dalal}},\ }\href@noop {} {\bibfield  {journal} {\bibinfo
  {journal} {Phys. Rev. Lett.}\ }\textbf {\bibinfo {volume} {96}},\ \bibinfo
  {pages} {107202} (\bibinfo {year} {2006})}\BibitemShut {NoStop}%
\bibitem [{\citenamefont {Yamase}\ \emph {et~al.}(2004)\citenamefont {Yamase},
  \citenamefont {Ishikawa}, \citenamefont {Fukaya}, \citenamefont {Nojiri},
  \citenamefont {Taniguchi},\ and\ \citenamefont {Atake}}]{Yamase2004}%
  \BibitemOpen
  \bibfield  {author} {\bibinfo {author} {\bibfnamefont {T.}~\bibnamefont
  {Yamase}}, \bibinfo {author} {\bibfnamefont {E.}~\bibnamefont {Ishikawa}},
  \bibinfo {author} {\bibfnamefont {K.}~\bibnamefont {Fukaya}}, \bibinfo
  {author} {\bibfnamefont {H.}~\bibnamefont {Nojiri}}, \bibinfo {author}
  {\bibfnamefont {T.}~\bibnamefont {Taniguchi}}, \ and\ \bibinfo {author}
  {\bibfnamefont {T.}~\bibnamefont {Atake}},\ }\href@noop {} {\bibfield
  {journal} {\bibinfo  {journal} {Inorg. Chem.}\ }\textbf {\bibinfo {volume}
  {43}},\ \bibinfo {pages} {8150} (\bibinfo {year} {2004})}\BibitemShut
  {NoStop}%
\bibitem [{\citenamefont {Trif}\ \emph {et~al.}(2010)\citenamefont {Trif},
  \citenamefont {Troiani}, \citenamefont {Stepanenko},\ and\ \citenamefont
  {Loss}}]{Trif2010}%
  \BibitemOpen
  \bibfield  {author} {\bibinfo {author} {\bibfnamefont {M.}~\bibnamefont
  {Trif}}, \bibinfo {author} {\bibfnamefont {F.}~\bibnamefont {Troiani}},
  \bibinfo {author} {\bibfnamefont {D.}~\bibnamefont {Stepanenko}}, \ and\
  \bibinfo {author} {\bibfnamefont {D.}~\bibnamefont {Loss}},\ }\href@noop {}
  {\bibfield  {journal} {\bibinfo  {journal} {Phys. Rev. B}\ }\textbf {\bibinfo
  {volume} {82}},\ \bibinfo {pages} {045429} (\bibinfo {year}
  {2010})}\BibitemShut {NoStop}%
\bibitem [{\citenamefont {Bulaevskii}\ \emph {et~al.}(2008)\citenamefont
  {Bulaevskii}, \citenamefont {Batista}, \citenamefont {Mostovoy},\ and\
  \citenamefont {Khomskii}}]{Bulaevskii2008}%
  \BibitemOpen
  \bibfield  {author} {\bibinfo {author} {\bibfnamefont {L.~N.}\ \bibnamefont
  {Bulaevskii}}, \bibinfo {author} {\bibfnamefont {C.~D.}\ \bibnamefont
  {Batista}}, \bibinfo {author} {\bibfnamefont {M.~V.}\ \bibnamefont
  {Mostovoy}}, \ and\ \bibinfo {author} {\bibfnamefont {D.~I.}\ \bibnamefont
  {Khomskii}},\ }\href {http://link.aps.org/doi/10.1103/PhysRevB.78.024402}
  {\bibfield  {journal} {\bibinfo  {journal} {Phys. Rev. B}\ }\textbf {\bibinfo
  {volume} {78}},\ \bibinfo {pages} {024402} (\bibinfo {year}
  {2008})}\BibitemShut {NoStop}%
\bibitem [{\citenamefont {Khomskii}(2010)}]{Khomskii2010}%
  \BibitemOpen
  \bibfield  {author} {\bibinfo {author} {\bibfnamefont {D.~I.}\ \bibnamefont
  {Khomskii}},\ }\href {http://stacks.iop.org/0953-8984/22/i=16/a=164209}
  {\bibfield  {journal} {\bibinfo  {journal} {Journal of Physics: Condensed
  Matter}\ }\textbf {\bibinfo {volume} {22}},\ \bibinfo {pages} {164209}
  (\bibinfo {year} {2010})}\BibitemShut {NoStop}%
\bibitem [{\citenamefont {Islam}\ \emph {et~al.}(2010)\citenamefont {Islam},
  \citenamefont {Nossa}, \citenamefont {Canali},\ and\ \citenamefont
  {Pederson}}]{Islam2010}%
  \BibitemOpen
  \bibfield  {author} {\bibinfo {author} {\bibfnamefont {M.~F.}\ \bibnamefont
  {Islam}}, \bibinfo {author} {\bibfnamefont {J.~F.}\ \bibnamefont {Nossa}},
  \bibinfo {author} {\bibfnamefont {C.~M.}\ \bibnamefont {Canali}}, \ and\
  \bibinfo {author} {\bibfnamefont {M.}~\bibnamefont {Pederson}},\ }\href@noop
  {} {\bibfield  {journal} {\bibinfo  {journal} {Phys. Rev. B}\ }\textbf
  {\bibinfo {volume} {82}},\ \bibinfo {pages} {155446} (\bibinfo {year}
  {2010})}\BibitemShut {NoStop}%
\bibitem [{\citenamefont {Khomskii}(2012)}]{Khomskii2012}%
  \BibitemOpen
  \bibfield  {author} {\bibinfo {author} {\bibfnamefont {D.}~\bibnamefont
  {Khomskii}},\ }\href {http://dx.doi.org/10.1038/ncomms1904} {\bibfield
  {journal} {\bibinfo  {journal} {Nat Commun}\ }\textbf {\bibinfo {volume}
  {3}},\ \bibinfo {pages} {904} (\bibinfo {year} {2012})}\BibitemShut {NoStop}%
\bibitem [{\citenamefont {Nossa}\ \emph {et~al.}(2013)\citenamefont {Nossa},
  \citenamefont {Islam}, \citenamefont {Canali},\ and\ \citenamefont
  {Pederson}}]{Nossa_SE_V15_2013}%
  \BibitemOpen
  \bibfield  {author} {\bibinfo {author} {\bibfnamefont {J.~F.}\ \bibnamefont
  {Nossa}}, \bibinfo {author} {\bibfnamefont {M.~F.}\ \bibnamefont {Islam}},
  \bibinfo {author} {\bibfnamefont {C.~M.}\ \bibnamefont {Canali}}, \ and\
  \bibinfo {author} {\bibfnamefont {M.~R.}\ \bibnamefont {Pederson}},\
  }\href@noop {} {\enquote {\bibinfo {title} {Electric control of spin states
  in frustrated triangular molecular magnets},}\ } (\bibinfo {year} {2013}),\
  \bibinfo {note} {unpublished}\BibitemShut {NoStop}%
\bibitem [{\citenamefont {van Hoogdalem}\ and\ \citenamefont
  {Loss}(2013)}]{Hoogdalem2013}%
  \BibitemOpen
  \bibfield  {author} {\bibinfo {author} {\bibfnamefont {K.~A.}\ \bibnamefont
  {van Hoogdalem}}\ and\ \bibinfo {author} {\bibfnamefont {D.}~\bibnamefont
  {Loss}},\ }\href {http://link.aps.org/doi/10.1103/PhysRevB.88.024420}
  {\bibfield  {journal} {\bibinfo  {journal} {Phys. Rev. B}\ }\textbf {\bibinfo
  {volume} {88}},\ \bibinfo {pages} {024420} (\bibinfo {year}
  {2013})}\BibitemShut {NoStop}%
\bibitem [{\citenamefont {Bulka}\ \emph {et~al.}(2011)\citenamefont {Bulka},
  \citenamefont {Kostyrko},\ and\ \citenamefont {Luczak}}]{Bulka2011}%
  \BibitemOpen
  \bibfield  {author} {\bibinfo {author} {\bibfnamefont {B.~R.}\ \bibnamefont
  {Bulka}}, \bibinfo {author} {\bibfnamefont {T.}~\bibnamefont {Kostyrko}}, \
  and\ \bibinfo {author} {\bibfnamefont {J.}~\bibnamefont {Luczak}},\ }\href
  {http://link.aps.org/doi/10.1103/PhysRevB.83.035301} {\bibfield  {journal}
  {\bibinfo  {journal} {Phys. Rev. B}\ }\textbf {\bibinfo {volume} {83}},\
  \bibinfo {pages} {035301} (\bibinfo {year} {2011})}\BibitemShut {NoStop}%
\bibitem [{\citenamefont {Weymann}\ \emph {et~al.}(2011)\citenamefont
  {Weymann}, \citenamefont {Bulka},\ and\ \citenamefont
  {Barnas}}]{Weymann2011}%
  \BibitemOpen
  \bibfield  {author} {\bibinfo {author} {\bibfnamefont {I.}~\bibnamefont
  {Weymann}}, \bibinfo {author} {\bibfnamefont {B.~R.}\ \bibnamefont {Bulka}},
  \ and\ \bibinfo {author} {\bibfnamefont {J.}~\bibnamefont {Barnas}},\ }\href
  {http://link.aps.org/doi/10.1103/PhysRevB.83.195302} {\bibfield  {journal}
  {\bibinfo  {journal} {Phys. Rev. B}\ }\textbf {\bibinfo {volume} {83}},\
  \bibinfo {pages} {195302} (\bibinfo {year} {2011})}\BibitemShut {NoStop}%
\bibitem [{\citenamefont {Luczak}\ and\ \citenamefont
  {Bulka}(2012)}]{Luczak2012}%
  \BibitemOpen
  \bibfield  {author} {\bibinfo {author} {\bibfnamefont {J.}~\bibnamefont
  {Luczak}}\ and\ \bibinfo {author} {\bibfnamefont {B.~R.}\ \bibnamefont
  {Bulka}},\ }\href {http://stacks.iop.org/0953-8984/24/i=37/a=375303}
  {\bibfield  {journal} {\bibinfo  {journal} {Journal of Physics: Condensed
  Matter}\ }\textbf {\bibinfo {volume} {24}},\ \bibinfo {pages} {375303}
  (\bibinfo {year} {2012})}\BibitemShut {NoStop}%
\bibitem [{\citenamefont {Hsieh}\ \emph {et~al.}(2012)\citenamefont {Hsieh},
  \citenamefont {Shim}, \citenamefont {Korkusinski},\ and\ \citenamefont
  {Hawrylak}}]{Hsieh2012}%
  \BibitemOpen
  \bibfield  {author} {\bibinfo {author} {\bibfnamefont {C.-Y.}\ \bibnamefont
  {Hsieh}}, \bibinfo {author} {\bibfnamefont {Y.-P.}\ \bibnamefont {Shim}},
  \bibinfo {author} {\bibfnamefont {M.}~\bibnamefont {Korkusinski}}, \ and\
  \bibinfo {author} {\bibfnamefont {P.}~\bibnamefont {Hawrylak}},\ }\href
  {http://stacks.iop.org/0034-4885/75/i=11/a=114501} {\bibfield  {journal}
  {\bibinfo  {journal} {Reports on Progress in Physics}\ }\textbf {\bibinfo
  {volume} {75}},\ \bibinfo {pages} {114501} (\bibinfo {year}
  {2012})}\BibitemShut {NoStop}%
\bibitem [{\citenamefont {Xiong}\ \emph {et~al.}(2012)\citenamefont {Xiong},
  \citenamefont {Huang},\ and\ \citenamefont {Wang}}]{Xiong2012}%
  \BibitemOpen
  \bibfield  {author} {\bibinfo {author} {\bibfnamefont {Y.-C.}\ \bibnamefont
  {Xiong}}, \bibinfo {author} {\bibfnamefont {J.}~\bibnamefont {Huang}}, \ and\
  \bibinfo {author} {\bibfnamefont {W.-Z.}\ \bibnamefont {Wang}},\ }\href
  {http://stacks.iop.org/0953-8984/24/i=45/a=455604} {\bibfield  {journal}
  {\bibinfo  {journal} {Journal of Physics: Condensed Matter}\ }\textbf
  {\bibinfo {volume} {24}},\ \bibinfo {pages} {455604} (\bibinfo {year}
  {2012})}\BibitemShut {NoStop}%
\bibitem [{\citenamefont {Friedel}\ \emph {et~al.}(1964)\citenamefont
  {Friedel}, \citenamefont {Lenglart},\ and\ \citenamefont
  {Leman}}]{friedel_1964}%
  \BibitemOpen
  \bibfield  {author} {\bibinfo {author} {\bibfnamefont {J.}~\bibnamefont
  {Friedel}}, \bibinfo {author} {\bibfnamefont {P.}~\bibnamefont {Lenglart}}, \
  and\ \bibinfo {author} {\bibfnamefont {G.}~\bibnamefont {Leman}},\
  }\href@noop {} {\bibfield  {journal} {\bibinfo  {journal} {J. Phys. Chem.
  Solids.}\ }\textbf {\bibinfo {volume} {25}},\ \bibinfo {pages} {781}
  (\bibinfo {year} {1964})}\BibitemShut {NoStop}%
\bibitem [{\citenamefont {Kaplan}(1983)}]{kaplan_1982}%
  \BibitemOpen
  \bibfield  {author} {\bibinfo {author} {\bibfnamefont {T.~A.}\ \bibnamefont
  {Kaplan}},\ }\href@noop {} {\bibfield  {journal} {\bibinfo  {journal} {Z.
  Phys. B - Condensed Matter}\ }\textbf {\bibinfo {volume} {49}},\ \bibinfo
  {pages} {313} (\bibinfo {year} {1983})}\BibitemShut {NoStop}%
\bibitem [{\citenamefont {Bonesteel}\ \emph {et~al.}(1992)\citenamefont
  {Bonesteel}, \citenamefont {Rice},\ and\ \citenamefont
  {Zhang}}]{bonesteel1992}%
  \BibitemOpen
  \bibfield  {author} {\bibinfo {author} {\bibfnamefont {N.~E.}\ \bibnamefont
  {Bonesteel}}, \bibinfo {author} {\bibfnamefont {T.~M.}\ \bibnamefont {Rice}},
  \ and\ \bibinfo {author} {\bibfnamefont {F.~C.}\ \bibnamefont {Zhang}},\
  }\href@noop {} {\bibfield  {journal} {\bibinfo  {journal} {Phys. Rev. Lett.}\
  }\textbf {\bibinfo {volume} {68}},\ \bibinfo {pages} {2684} (\bibinfo {year}
  {1992})}\BibitemShut {NoStop}%
\bibitem [{\citenamefont {Moriya}(1960)}]{Moriya1960}%
  \BibitemOpen
  \bibfield  {author} {\bibinfo {author} {\bibfnamefont {T.}~\bibnamefont
  {Moriya}},\ }\href@noop {} {\bibfield  {journal} {\bibinfo  {journal} {Phys.
  Rev.}\ }\textbf {\bibinfo {volume} {120}},\ \bibinfo {pages} {91} (\bibinfo
  {year} {1960})}\BibitemShut {NoStop}%
\bibitem [{\citenamefont {Nossa}\ \emph {et~al.}(2012)\citenamefont {Nossa},
  \citenamefont {Islam}, \citenamefont {Canali},\ and\ \citenamefont
  {Pederson}}]{Nossa2012}%
  \BibitemOpen
  \bibfield  {author} {\bibinfo {author} {\bibfnamefont {J.~F.}\ \bibnamefont
  {Nossa}}, \bibinfo {author} {\bibfnamefont {M.~F.}\ \bibnamefont {Islam}},
  \bibinfo {author} {\bibfnamefont {C.~M.}\ \bibnamefont {Canali}}, \ and\
  \bibinfo {author} {\bibfnamefont {M.~R.}\ \bibnamefont {Pederson}},\ }\href
  {http://link.aps.org/doi/10.1103/PhysRevB.85.085427} {\bibfield  {journal}
  {\bibinfo  {journal} {Phys. Rev. B}\ }\textbf {\bibinfo {volume} {85}},\
  \bibinfo {pages} {085427} (\bibinfo {year} {2012})}\BibitemShut {NoStop}%
\bibitem [{\citenamefont {Boris}(2006)}]{Boris2006}%
  \BibitemOpen
  \bibfield  {author} {\bibinfo {author} {\bibfnamefont {S.~T.}\ \bibnamefont
  {Boris}},\ }\href@noop {} {\emph {\bibinfo {title} {Group Theory in Chemistry
  and Spectroscopy.}}}\ (\bibinfo  {publisher} {Dover publications, INC},\
  \bibinfo {year} {2006})\BibitemShut {NoStop}%
\bibitem [{\citenamefont {Fuechsle}\ \emph {et~al.}(2012)\citenamefont
  {Fuechsle}, \citenamefont {Miwa}, \citenamefont {Mahapatra}, \citenamefont
  {Ryu}, \citenamefont {Lee}, \citenamefont {Warschkow}, \citenamefont
  {Hollenberg}, \citenamefont {Klimeck},\ and\ \citenamefont
  {Simmons}}]{Fuechsle2012}%
  \BibitemOpen
  \bibfield  {author} {\bibinfo {author} {\bibfnamefont {M.}~\bibnamefont
  {Fuechsle}}, \bibinfo {author} {\bibfnamefont {J.~A.}\ \bibnamefont {Miwa}},
  \bibinfo {author} {\bibfnamefont {S.}~\bibnamefont {Mahapatra}}, \bibinfo
  {author} {\bibfnamefont {H.}~\bibnamefont {Ryu}}, \bibinfo {author}
  {\bibfnamefont {S.}~\bibnamefont {Lee}}, \bibinfo {author} {\bibfnamefont
  {O.}~\bibnamefont {Warschkow}}, \bibinfo {author} {\bibfnamefont {L.~C.~L.}\
  \bibnamefont {Hollenberg}}, \bibinfo {author} {\bibfnamefont
  {G.}~\bibnamefont {Klimeck}}, \ and\ \bibinfo {author} {\bibfnamefont
  {M.~Y.}\ \bibnamefont {Simmons}},\ }\href@noop {} {\bibfield  {journal}
  {\bibinfo  {journal} {Nature Nanotechnology}\ }\textbf {\bibinfo {volume}
  {7}},\ \bibinfo {pages} {242} (\bibinfo {year} {2012})}\BibitemShut {NoStop}%
\bibitem [{\citenamefont {Heersche}\ \emph {et~al.}(2006)\citenamefont
  {Heersche}, \citenamefont {de~Groot}, \citenamefont {Folk}, \citenamefont
  {van~der Zant}, \citenamefont {Romeike}, \citenamefont {Wegewijs},
  \citenamefont {Zobbi}, \citenamefont {Barreca}, \citenamefont {Tondello},\
  and\ \citenamefont {Cornia}}]{Heersche2006}%
  \BibitemOpen
  \bibfield  {author} {\bibinfo {author} {\bibfnamefont {H.~B.}\ \bibnamefont
  {Heersche}}, \bibinfo {author} {\bibfnamefont {Z.}~\bibnamefont {de~Groot}},
  \bibinfo {author} {\bibfnamefont {J.~A.}\ \bibnamefont {Folk}}, \bibinfo
  {author} {\bibfnamefont {H.~S.~J.}\ \bibnamefont {van~der Zant}}, \bibinfo
  {author} {\bibfnamefont {C.}~\bibnamefont {Romeike}}, \bibinfo {author}
  {\bibfnamefont {M.~R.}\ \bibnamefont {Wegewijs}}, \bibinfo {author}
  {\bibfnamefont {L.}~\bibnamefont {Zobbi}}, \bibinfo {author} {\bibfnamefont
  {D.}~\bibnamefont {Barreca}}, \bibinfo {author} {\bibfnamefont
  {E.}~\bibnamefont {Tondello}}, \ and\ \bibinfo {author} {\bibfnamefont
  {A.}~\bibnamefont {Cornia}},\ }\href
  {http://link.aps.org/doi/10.1103/PhysRevLett.96.206801} {\bibfield  {journal}
  {\bibinfo  {journal} {Phys. Rev. Lett.}\ }\textbf {\bibinfo {volume} {96}},\
  \bibinfo {pages} {206801} (\bibinfo {year} {2006})}\BibitemShut {NoStop}%
\bibitem [{\citenamefont {Jo}\ \emph {et~al.}(2006)\citenamefont {Jo},
  \citenamefont {Grose}, \citenamefont {Baheti}, \citenamefont {Deshmukh},
  \citenamefont {Sokol}, \citenamefont {Rumberger}, \citenamefont
  {Hendrickson}, \citenamefont {Long}, \citenamefont {Park},\ and\
  \citenamefont {Ralph}}]{Jo2006}%
  \BibitemOpen
  \bibfield  {author} {\bibinfo {author} {\bibfnamefont {M.-H.}\ \bibnamefont
  {Jo}}, \bibinfo {author} {\bibfnamefont {J.~E.}\ \bibnamefont {Grose}},
  \bibinfo {author} {\bibfnamefont {K.}~\bibnamefont {Baheti}}, \bibinfo
  {author} {\bibfnamefont {M.~M.}\ \bibnamefont {Deshmukh}}, \bibinfo {author}
  {\bibfnamefont {J.~J.}\ \bibnamefont {Sokol}}, \bibinfo {author}
  {\bibfnamefont {E.~M.}\ \bibnamefont {Rumberger}}, \bibinfo {author}
  {\bibfnamefont {D.~N.}\ \bibnamefont {Hendrickson}}, \bibinfo {author}
  {\bibfnamefont {J.~R.}\ \bibnamefont {Long}}, \bibinfo {author}
  {\bibfnamefont {H.}~\bibnamefont {Park}}, \ and\ \bibinfo {author}
  {\bibfnamefont {D.~C.}\ \bibnamefont {Ralph}},\ }\bibfield  {booktitle}
  {\emph {\bibinfo {booktitle} {Nano Letters}},\ }\href {\doibase
  10.1021/nl061212i} {\bibfield  {journal} {\bibinfo  {journal} {Nano Lett.}\
  }\textbf {\bibinfo {volume} {6}},\ \bibinfo {pages} {2014} (\bibinfo {year}
  {2006})}\BibitemShut {NoStop}%
\bibitem [{\citenamefont {Elste}\ and\ \citenamefont {Timm}(2005)}]{Elste2005}%
  \BibitemOpen
  \bibfield  {author} {\bibinfo {author} {\bibfnamefont {F.}~\bibnamefont
  {Elste}}\ and\ \bibinfo {author} {\bibfnamefont {C.}~\bibnamefont {Timm}},\
  }\href@noop {} {\bibfield  {journal} {\bibinfo  {journal} {Phys. Rev. B}\
  }\textbf {\bibinfo {volume} {71}},\ \bibinfo {pages} {155403} (\bibinfo
  {year} {2005})}\BibitemShut {NoStop}%
\bibitem [{\citenamefont {Leijnse}\ and\ \citenamefont
  {Wegewijs}(2008)}]{Leijnse2008}%
  \BibitemOpen
  \bibfield  {author} {\bibinfo {author} {\bibfnamefont {M.}~\bibnamefont
  {Leijnse}}\ and\ \bibinfo {author} {\bibfnamefont {M.~R.}\ \bibnamefont
  {Wegewijs}},\ }\href {http://link.aps.org/doi/10.1103/PhysRevB.78.235424}
  {\bibfield  {journal} {\bibinfo  {journal} {Phys. Rev. B}\ }\textbf {\bibinfo
  {volume} {78}},\ \bibinfo {pages} {235424} (\bibinfo {year}
  {2008})}\BibitemShut {NoStop}%
\bibitem [{\citenamefont {Koller}\ \emph {et~al.}(2010)\citenamefont {Koller},
  \citenamefont {Grifoni}, \citenamefont {Leijnse},\ and\ \citenamefont
  {Wegewijs}}]{Koller2010}%
  \BibitemOpen
  \bibfield  {author} {\bibinfo {author} {\bibfnamefont {S.}~\bibnamefont
  {Koller}}, \bibinfo {author} {\bibfnamefont {M.}~\bibnamefont {Grifoni}},
  \bibinfo {author} {\bibfnamefont {M.}~\bibnamefont {Leijnse}}, \ and\
  \bibinfo {author} {\bibfnamefont {M.~R.}\ \bibnamefont {Wegewijs}},\ }\href
  {http://link.aps.org/doi/10.1103/PhysRevB.82.235307} {\bibfield  {journal}
  {\bibinfo  {journal} {Phys. Rev. B}\ }\textbf {\bibinfo {volume} {82}},\
  \bibinfo {pages} {235307} (\bibinfo {year} {2010})}\BibitemShut {NoStop}%
\bibitem [{\citenamefont {Weinmann}\ \emph {et~al.}(1995)\citenamefont
  {Weinmann}, \citenamefont {Häusler},\ and\ \citenamefont
  {Kramer}}]{Weinmann1995}%
  \BibitemOpen
  \bibfield  {author} {\bibinfo {author} {\bibfnamefont {D.}~\bibnamefont
  {Weinmann}}, \bibinfo {author} {\bibfnamefont {W.}~\bibnamefont {Häusler}},
  \ and\ \bibinfo {author} {\bibfnamefont {B.}~\bibnamefont {Kramer}},\ }\href
  {http://link.aps.org/doi/10.1103/PhysRevLett.74.984} {\bibfield  {journal}
  {\bibinfo  {journal} {Phys. Rev. Lett.}\ }\textbf {\bibinfo {volume} {74}},\
  \bibinfo {pages} {984} (\bibinfo {year} {1995})}\BibitemShut {NoStop}%
\bibitem [{\citenamefont {Leijnse}\ \emph {et~al.}(2009)\citenamefont
  {Leijnse}, \citenamefont {Wegewijs},\ and\ \citenamefont
  {Hettler}}]{Leijnse2009}%
  \BibitemOpen
  \bibfield  {author} {\bibinfo {author} {\bibfnamefont {M.}~\bibnamefont
  {Leijnse}}, \bibinfo {author} {\bibfnamefont {M.~R.}\ \bibnamefont
  {Wegewijs}}, \ and\ \bibinfo {author} {\bibfnamefont {M.~H.}\ \bibnamefont
  {Hettler}},\ }\href {http://link.aps.org/doi/10.1103/PhysRevLett.103.156803}
  {\bibfield  {journal} {\bibinfo  {journal} {Phys. Rev. Lett.}\ }\textbf
  {\bibinfo {volume} {103}},\ \bibinfo {pages} {156803} (\bibinfo {year}
  {2009})}\BibitemShut {NoStop}%
\bibitem [{\citenamefont {Timm}\ and\ \citenamefont
  {Elste}(2006)}]{timm2006prb}%
  \BibitemOpen
  \bibfield  {author} {\bibinfo {author} {\bibfnamefont {C.}~\bibnamefont
  {Timm}}\ and\ \bibinfo {author} {\bibfnamefont {F.}~\bibnamefont {Elste}},\
  }\href@noop {} {\bibfield  {journal} {\bibinfo  {journal} {Phys. Rev. B}\
  }\textbf {\bibinfo {volume} {73}},\ \bibinfo {pages} {235304} (\bibinfo
  {year} {2006})}\BibitemShut {NoStop}%
\bibitem [{\citenamefont {Elste}\ and\ \citenamefont
  {Timm}(2006)}]{elste2006prb}%
  \BibitemOpen
  \bibfield  {author} {\bibinfo {author} {\bibfnamefont {F.}~\bibnamefont
  {Elste}}\ and\ \bibinfo {author} {\bibfnamefont {C.}~\bibnamefont {Timm}},\
  }\href@noop {} {\bibfield  {journal} {\bibinfo  {journal} {Phys. Rev. B}\
  }\textbf {\bibinfo {volume} {73}},\ \bibinfo {pages} {235305} (\bibinfo
  {year} {2006})}\BibitemShut {NoStop}%
\bibitem [{\citenamefont {Elste}\ and\ \citenamefont
  {Timm}(2007)}]{elste2007prb}%
  \BibitemOpen
  \bibfield  {author} {\bibinfo {author} {\bibfnamefont {F.}~\bibnamefont
  {Elste}}\ and\ \bibinfo {author} {\bibfnamefont {C.}~\bibnamefont {Timm}},\
  }\href@noop {} {\bibfield  {journal} {\bibinfo  {journal} {Phys. Rev. B}\
  }\textbf {\bibinfo {volume} {75}},\ \bibinfo {pages} {195341} (\bibinfo
  {year} {2007})}\BibitemShut {NoStop}%
\bibitem [{\citenamefont {Timm}(2007)}]{timm2007prb}%
  \BibitemOpen
  \bibfield  {author} {\bibinfo {author} {\bibfnamefont {C.}~\bibnamefont
  {Timm}},\ }\href@noop {} {\bibfield  {journal} {\bibinfo  {journal} {Phys.
  Rev. B}\ }\textbf {\bibinfo {volume} {76}},\ \bibinfo {pages} {014421}
  (\bibinfo {year} {2007})}\BibitemShut {NoStop}%
\bibitem [{\citenamefont {Bruus}\ and\ \citenamefont
  {Flensberg}(2004)}]{Bruus2004}%
  \BibitemOpen
  \bibfield  {author} {\bibinfo {author} {\bibfnamefont {H.}~\bibnamefont
  {Bruus}}\ and\ \bibinfo {author} {\bibfnamefont {K.}~\bibnamefont
  {Flensberg}},\ }\href@noop {} {\emph {\bibinfo {title} {Many body quantum
  theory in condensed matter physics}}}\ (\bibinfo  {publisher} {Oxford
  Graduate Texts},\ \bibinfo {year} {2004})\BibitemShut {NoStop}%
\bibitem [{\citenamefont {Tews}(2004)}]{Tews2004}%
  \BibitemOpen
  \bibfield  {author} {\bibinfo {author} {\bibfnamefont {M.}~\bibnamefont
  {Tews}},\ }\href {\doibase 10.1002/andp.200410076} {\bibfield  {journal}
  {\bibinfo  {journal} {Annalen der Physik}\ }\textbf {\bibinfo {volume}
  {13}},\ \bibinfo {pages} {249} (\bibinfo {year} {2004})}\BibitemShut
  {NoStop}%
\bibitem [{\citenamefont {Koch}\ \emph {et~al.}(2006)\citenamefont {Koch},
  \citenamefont {Raikh},\ and\ \citenamefont {von Oppen}}]{Koch2006}%
  \BibitemOpen
  \bibfield  {author} {\bibinfo {author} {\bibfnamefont {J.}~\bibnamefont
  {Koch}}, \bibinfo {author} {\bibfnamefont {M.~E.}\ \bibnamefont {Raikh}}, \
  and\ \bibinfo {author} {\bibfnamefont {F.}~\bibnamefont {von Oppen}},\ }\href
  {http://link.aps.org/doi/10.1103/PhysRevLett.96.056803} {\bibfield  {journal}
  {\bibinfo  {journal} {Phys. Rev. Lett.}\ }\textbf {\bibinfo {volume} {96}},\
  \bibinfo {pages} {056803} (\bibinfo {year} {2006})}\BibitemShut {NoStop}%
\bibitem [{\citenamefont {Turek}\ and\ \citenamefont
  {Matveev}(2002)}]{Turek2002}%
  \BibitemOpen
  \bibfield  {author} {\bibinfo {author} {\bibfnamefont {M.}~\bibnamefont
  {Turek}}\ and\ \bibinfo {author} {\bibfnamefont {K.~A.}\ \bibnamefont
  {Matveev}},\ }\href {http://link.aps.org/doi/10.1103/PhysRevB.65.115332}
  {\bibfield  {journal} {\bibinfo  {journal} {Phys. Rev. B}\ }\textbf {\bibinfo
  {volume} {65}},\ \bibinfo {pages} {115332} (\bibinfo {year}
  {2002})}\BibitemShut {NoStop}%
\bibitem [{\citenamefont {Koch}\ \emph {et~al.}(2004)\citenamefont {Koch},
  \citenamefont {von Oppen}, \citenamefont {Oreg},\ and\ \citenamefont
  {Sela}}]{Koch2004}%
  \BibitemOpen
  \bibfield  {author} {\bibinfo {author} {\bibfnamefont {J.}~\bibnamefont
  {Koch}}, \bibinfo {author} {\bibfnamefont {F.}~\bibnamefont {von Oppen}},
  \bibinfo {author} {\bibfnamefont {Y.}~\bibnamefont {Oreg}}, \ and\ \bibinfo
  {author} {\bibfnamefont {E.}~\bibnamefont {Sela}},\ }\href
  {http://link.aps.org/doi/10.1103/PhysRevB.70.195107} {\bibfield  {journal}
  {\bibinfo  {journal} {Phys. Rev. B}\ }\textbf {\bibinfo {volume} {70}},\
  \bibinfo {pages} {195107} (\bibinfo {year} {2004})}\BibitemShut {NoStop}%
\bibitem [{\citenamefont {Schleser}\ \emph {et~al.}(2005)\citenamefont
  {Schleser}, \citenamefont {Ihn}, \citenamefont {Ruh}, \citenamefont
  {Ensslin}, \citenamefont {Tews}, \citenamefont {Pfannkuche}, \citenamefont
  {Driscoll},\ and\ \citenamefont {Gossard}}]{Schleser2005}%
  \BibitemOpen
  \bibfield  {author} {\bibinfo {author} {\bibfnamefont {R.}~\bibnamefont
  {Schleser}}, \bibinfo {author} {\bibfnamefont {T.}~\bibnamefont {Ihn}},
  \bibinfo {author} {\bibfnamefont {E.}~\bibnamefont {Ruh}}, \bibinfo {author}
  {\bibfnamefont {K.}~\bibnamefont {Ensslin}}, \bibinfo {author} {\bibfnamefont
  {M.}~\bibnamefont {Tews}}, \bibinfo {author} {\bibfnamefont {D.}~\bibnamefont
  {Pfannkuche}}, \bibinfo {author} {\bibfnamefont {D.~C.}\ \bibnamefont
  {Driscoll}}, \ and\ \bibinfo {author} {\bibfnamefont {A.~C.}\ \bibnamefont
  {Gossard}},\ }\href {\doibase 10.1103/PhysRevLett.94.206805} {\bibfield
  {journal} {\bibinfo  {journal} {Phys. Rev. Lett.}\ }\textbf {\bibinfo
  {volume} {94}},\ \bibinfo {pages} {206805} (\bibinfo {year}
  {2005})}\BibitemShut {NoStop}%
\bibitem [{\citenamefont {Holm}\ \emph {et~al.}(2008)\citenamefont {Holm},
  \citenamefont {J\o{}rgensen}, \citenamefont {Grove-Rasmussen}, \citenamefont
  {Paaske}, \citenamefont {Flensberg},\ and\ \citenamefont
  {Lindelof}}]{Holm2008}%
  \BibitemOpen
  \bibfield  {author} {\bibinfo {author} {\bibfnamefont {J.~V.}\ \bibnamefont
  {Holm}}, \bibinfo {author} {\bibfnamefont {H.~I.}\ \bibnamefont
  {J\o{}rgensen}}, \bibinfo {author} {\bibfnamefont {K.}~\bibnamefont
  {Grove-Rasmussen}}, \bibinfo {author} {\bibfnamefont {J.}~\bibnamefont
  {Paaske}}, \bibinfo {author} {\bibfnamefont {K.}~\bibnamefont {Flensberg}}, \
  and\ \bibinfo {author} {\bibfnamefont {P.~E.}\ \bibnamefont {Lindelof}},\
  }\href {\doibase 10.1103/PhysRevB.77.161406} {\bibfield  {journal} {\bibinfo
  {journal} {Phys. Rev. B}\ }\textbf {\bibinfo {volume} {77}},\ \bibinfo
  {pages} {161406} (\bibinfo {year} {2008})}\BibitemShut {NoStop}%
\bibitem [{\citenamefont {Paaske}\ \emph {et~al.}(2006)\citenamefont {Paaske},
  \citenamefont {Rosch}, \citenamefont {Wolfle}, \citenamefont {Mason},
  \citenamefont {Marcus},\ and\ \citenamefont {Nygard}}]{Paaske2006}%
  \BibitemOpen
  \bibfield  {author} {\bibinfo {author} {\bibfnamefont {J.}~\bibnamefont
  {Paaske}}, \bibinfo {author} {\bibfnamefont {A.}~\bibnamefont {Rosch}},
  \bibinfo {author} {\bibfnamefont {P.}~\bibnamefont {Wolfle}}, \bibinfo
  {author} {\bibfnamefont {N.}~\bibnamefont {Mason}}, \bibinfo {author}
  {\bibfnamefont {C.~M.}\ \bibnamefont {Marcus}}, \ and\ \bibinfo {author}
  {\bibfnamefont {J.}~\bibnamefont {Nygard}},\ }\href
  {http://dx.doi.org/10.1038/nphys340} {\bibfield  {journal} {\bibinfo
  {journal} {Nat Phys}\ }\textbf {\bibinfo {volume} {2}},\ \bibinfo {pages}
  {460} (\bibinfo {year} {2006})}\BibitemShut {NoStop}%
\bibitem [{\citenamefont {Sapmaz}\ \emph {et~al.}(2005)\citenamefont {Sapmaz},
  \citenamefont {Jarillo-Herrero}, \citenamefont {Kong}, \citenamefont
  {Dekker}, \citenamefont {Kouwenhoven},\ and\ \citenamefont {van~der
  Zant}}]{Sapmaz2005}%
  \BibitemOpen
  \bibfield  {author} {\bibinfo {author} {\bibfnamefont {S.}~\bibnamefont
  {Sapmaz}}, \bibinfo {author} {\bibfnamefont {P.}~\bibnamefont
  {Jarillo-Herrero}}, \bibinfo {author} {\bibfnamefont {J.}~\bibnamefont
  {Kong}}, \bibinfo {author} {\bibfnamefont {C.}~\bibnamefont {Dekker}},
  \bibinfo {author} {\bibfnamefont {L.~P.}\ \bibnamefont {Kouwenhoven}}, \ and\
  \bibinfo {author} {\bibfnamefont {H.~S.~J.}\ \bibnamefont {van~der Zant}},\
  }\href {\doibase 10.1103/PhysRevB.71.153402} {\bibfield  {journal} {\bibinfo
  {journal} {Phys. Rev. B}\ }\textbf {\bibinfo {volume} {71}},\ \bibinfo
  {pages} {153402} (\bibinfo {year} {2005})}\BibitemShut {NoStop}%
\bibitem [{\citenamefont {Roch}\ \emph {et~al.}(2008)\citenamefont {Roch},
  \citenamefont {Florens}, \citenamefont {Bouchiat}, \citenamefont
  {Wernsdorfer},\ and\ \citenamefont {Balestro}}]{Roch2008}%
  \BibitemOpen
  \bibfield  {author} {\bibinfo {author} {\bibfnamefont {N.}~\bibnamefont
  {Roch}}, \bibinfo {author} {\bibfnamefont {S.}~\bibnamefont {Florens}},
  \bibinfo {author} {\bibfnamefont {V.}~\bibnamefont {Bouchiat}}, \bibinfo
  {author} {\bibfnamefont {W.}~\bibnamefont {Wernsdorfer}}, \ and\ \bibinfo
  {author} {\bibfnamefont {F.}~\bibnamefont {Balestro}},\ }\href
  {http://dx.doi.org/10.1038/nature06930} {\bibfield  {journal} {\bibinfo
  {journal} {Nature}\ }\textbf {\bibinfo {volume} {453}},\ \bibinfo {pages}
  {633} (\bibinfo {year} {2008})}\BibitemShut {NoStop}%
\bibitem [{\citenamefont {Parks}\ \emph {et~al.}(2007)\citenamefont {Parks},
  \citenamefont {Champagne}, \citenamefont {Hutchison}, \citenamefont
  {Flores-Torres}, \citenamefont {Abru\~na},\ and\ \citenamefont
  {Ralph}}]{Parks2007}%
  \BibitemOpen
  \bibfield  {author} {\bibinfo {author} {\bibfnamefont {J.~J.}\ \bibnamefont
  {Parks}}, \bibinfo {author} {\bibfnamefont {A.~R.}\ \bibnamefont
  {Champagne}}, \bibinfo {author} {\bibfnamefont {G.~R.}\ \bibnamefont
  {Hutchison}}, \bibinfo {author} {\bibfnamefont {S.}~\bibnamefont
  {Flores-Torres}}, \bibinfo {author} {\bibfnamefont {H.~D.}\ \bibnamefont
  {Abru\~na}}, \ and\ \bibinfo {author} {\bibfnamefont {D.~C.}\ \bibnamefont
  {Ralph}},\ }\href {\doibase 10.1103/PhysRevLett.99.026601} {\bibfield
  {journal} {\bibinfo  {journal} {Phys. Rev. Lett.}\ }\textbf {\bibinfo
  {volume} {99}},\ \bibinfo {pages} {026601} (\bibinfo {year}
  {2007})}\BibitemShut {NoStop}%
\bibitem [{\citenamefont {Osorio}\ \emph {et~al.}(2007)\citenamefont {Osorio},
  \citenamefont {O'Neill}, \citenamefont {Wegewijs}, \citenamefont
  {Stuhr-Hansen}, \citenamefont {Paaske}, \citenamefont {Bjørnholm},\ and\
  \citenamefont {van~der Zant}}]{Osorio2007}%
  \BibitemOpen
  \bibfield  {author} {\bibinfo {author} {\bibfnamefont {E.~A.}\ \bibnamefont
  {Osorio}}, \bibinfo {author} {\bibfnamefont {K.}~\bibnamefont {O'Neill}},
  \bibinfo {author} {\bibfnamefont {M.}~\bibnamefont {Wegewijs}}, \bibinfo
  {author} {\bibfnamefont {N.}~\bibnamefont {Stuhr-Hansen}}, \bibinfo {author}
  {\bibfnamefont {J.}~\bibnamefont {Paaske}}, \bibinfo {author} {\bibfnamefont
  {T.}~\bibnamefont {Bjørnholm}}, \ and\ \bibinfo {author} {\bibfnamefont
  {H.~S.~J.}\ \bibnamefont {van~der Zant}},\ }\bibfield  {booktitle} {\emph
  {\bibinfo {booktitle} {Nano Letters}},\ }\href {\doibase 10.1021/nl0715802}
  {\bibfield  {journal} {\bibinfo  {journal} {Nano Lett.}\ }\textbf {\bibinfo
  {volume} {7}},\ \bibinfo {pages} {3336} (\bibinfo {year} {2007})}\BibitemShut
  {NoStop}%
\bibitem [{\citenamefont {Galperin}\ \emph {et~al.}(2007)\citenamefont
  {Galperin}, \citenamefont {Ratner},\ and\ \citenamefont
  {Nitzan}}]{Galperin2007}%
  \BibitemOpen
  \bibfield  {author} {\bibinfo {author} {\bibfnamefont {M.}~\bibnamefont
  {Galperin}}, \bibinfo {author} {\bibfnamefont {M.~A.}\ \bibnamefont
  {Ratner}}, \ and\ \bibinfo {author} {\bibfnamefont {A.}~\bibnamefont
  {Nitzan}},\ }\href@noop {} {\bibfield  {journal} {\bibinfo  {journal} {J.
  Phys.: Condens. Matter}\ }\textbf {\bibinfo {volume} {19}},\ \bibinfo {pages}
  {103201} (\bibinfo {year} {2007})}\BibitemShut {NoStop}%
\bibitem [{\citenamefont {Reed}(2008)}]{Reed2008}%
  \BibitemOpen
  \bibfield  {author} {\bibinfo {author} {\bibfnamefont {M.~A.}\ \bibnamefont
  {Reed}},\ }\href@noop {} {\bibfield  {journal} {\bibinfo  {journal}
  {Materials Today}\ }\textbf {\bibinfo {volume} {11}},\ \bibinfo {pages} {46}
  (\bibinfo {year} {2008})}\BibitemShut {NoStop}%
\bibitem [{Note1()}]{Note1}%
  \BibitemOpen
  \bibinfo {note} {In principle, because of the presence of the spin-orbit
  interaction, states with different total $S$ are coupled. However the mixing
  is of the order of the Dzyaloshinskii-Moriya (DM) parameter $D \propto t
  \lambda _{\protect \rm SOI}/U$, which, for the parameters used here, is very
  small on the scale of the exchange constant separating states with different
  $S$. Therefore, in practice, $S$ and $S_z$ are good quantum
  numbers.}\BibitemShut {Stop}%
\bibitem [{\citenamefont {Mingo}\ and\ \citenamefont
  {Flores}(1998)}]{Mingo1998}%
  \BibitemOpen
  \bibfield  {author} {\bibinfo {author} {\bibfnamefont {N.}~\bibnamefont
  {Mingo}}\ and\ \bibinfo {author} {\bibfnamefont {F.}~\bibnamefont {Flores}},\
  }\href {http://www.sciencedirect.com/science/article/pii/S0040609097011413}
  {\bibfield  {journal} {\bibinfo  {journal} {Thin Solid Films}\ }\textbf
  {\bibinfo {volume} {318}},\ \bibinfo {pages} {69} (\bibinfo {year}
  {1998})}\BibitemShut {NoStop}%
\bibitem [{\citenamefont {Stokbro}\ \emph {et~al.}(1998)\citenamefont
  {Stokbro}, \citenamefont {Quaade},\ and\ \citenamefont {Grey}}]{Stokbro1998}%
  \BibitemOpen
  \bibfield  {author} {\bibinfo {author} {\bibfnamefont {K.}~\bibnamefont
  {Stokbro}}, \bibinfo {author} {\bibfnamefont {U.}~\bibnamefont {Quaade}}, \
  and\ \bibinfo {author} {\bibfnamefont {F.}~\bibnamefont {Grey}},\ }\href
  {http://dx.doi.org/10.1007/s003390051265} {\bibfield  {journal} {\bibinfo
  {journal} {Applied Physics A}\ }\textbf {\bibinfo {volume} {66}},\ \bibinfo
  {pages} {S907} (\bibinfo {year} {1998})}\BibitemShut {NoStop}%
\end{thebibliography}%

\end{document}